\documentclass{article}
\usepackage[utf8]{inputenc}
\usepackage{authblk}

\usepackage[english]{babel}
\usepackage{latexsym}
\usepackage{indentfirst,cite}
\usepackage{amsmath}
\usepackage{amssymb}
  \usepackage{pdfpages}
 \usepackage{bm}
 \usepackage{comment}
\usepackage{graphicx} 
\usepackage{amsfonts,soul,mathrsfs}

\newcommand{\be}{\begin{equation}}
 \newcommand{\ee}{\end{equation}}
\newcommand{\bear}{\be\begin{array}}
\newcommand{\bea}{\begin{eqnarray}}
\newcommand{\eea}{\end{eqnarray}}

\newcommand{\bp}{{\bf p}}
\newcommand{\br}{{\bf r}}

\newcommand{\bk}{{\bf k}}

\newcommand{\la}{\langle}
\newcommand{\ra}{\rangle}

\newcommand{\dst}{\displaystyle}
\newcommand{\fr}[2]{\frac{{\dst #1}}{{\dst #2}}}

\begin{document}

\title{Shifting physics of vortex particles to higher energies \\ via quantum entanglement
}

\author[$1$]{D.\,V.~Karlovets\thanks{Corresponding author}}
\author[$1$]{S.\,S.~Baturin}
\author[$2$]{G.~Geloni}
\author[$1$]{G.\,K.~Sizykh}
\author[$1,3,4$]{V.\,G.~Serbo}
\affil[$1$]{{\small School of Physics and Engineering, ITMO University, 
197101 St.\,Petersburg, Russia}}
\affil[$2$]{{\small European XFEL, Holzkoppel 4, 22869 Schenefeld, Germany}}
\affil[$3$]{{\small Novosibirsk State University, 630090 Novosibirsk, Russia}}
\affil[$4$]{{\small Sobolev Institute of Mathematics, 630090 Novosibirsk, Russia}}

\date{\today}


\maketitle

\begin{abstract}
Physics of structured waves is currently limited to relatively small particle energies as the available generation techniques are only applicable to the soft $X$-ray twisted photons, to the beams of electron microscopes, to cold neutrons, or non-relativistic atoms. The highly energetic vortex particles with an orbital angular momentum would come in handy for a number of experiments in atomic physics, nuclear, hadronic, and accelerator physics, and to generate them one needs to develop alternative methods, applicable for ultrarelativistic energies and for composite particles. Here, we show that the vortex states of in principle arbitrary particles can be generated during photon emission in helical undulators, via Cherenkov radiation, in collisions of charged particles with intense laser beams, in such scattering or annihilation processes as $e\mu \to e\mu, ep \to ep, e^-e^+ \to p\bar{p}$, and so forth. The key element in obtaining them is the postselection protocol due to entanglement between a pair of final particles and it is largely not the process itself. The state of a final particle -- be it a $\gamma$-ray, a hadron, a nucleus, or an ion -- becomes twisted if the azimuthal angle of the other particle momentum is measured with a large error or is not measured at all. As a result, requirements to the beam transverse coherence can be greatly relaxed, which enables the generation of highly energetic vortex beams at accelerators and synchrotron radiation facilities, thus making them a new tool for hadronic and spin studies. 
\end{abstract}


\section{Introduction}

Photons with an orbital angular momentum (OAM) \cite{Allen} have found applications in quantum optics and information, optomechanics, biology, astrophysics, and in other fields \cite{Torres, And, SerboUFN, Tamb, Budker, TairaX, Ivanov-atoms, Kick}. They can be entangled in the polarization and orbital degrees of freedom \cite{Mair, Bhatti} and their OAM quanta can reach the values of 10010$\hbar$\cite{Zeilinger10010}. Along with the diffraction techniques, such twisted light can be generated by charged particles in helical undulators \cite{Sasaki1, Sasaki2, Afan, Bahrdt, Kaneyasu, BKL, BKL2}, via non-linear Thomson or Compton scattering \cite{BKL2, Serbo2011-1, Serbo2011-2, Taira, Katoh} or -- more generally -- when the electron trajectory is helical \cite{Katoh2, Epp, Epp2}, during Cherenkov or transition radiation \cite{BKLCh, BKLKh}, via channeling in crystals \cite{Chan1, Chan2}, and so forth. However, despite the proposals to use twisted photons in particle and nuclear studies \cite{Budker}, their highest energy achieved so far does not exceed but a few keV \cite{TairaX}. One recent idea to upconvert their frequency is to employ resonant scattering of the optical twisted photons on relativistic ions within the Gamma Factory project at the Large Hadron Collider \cite{Budker, PRA21}.

The massive counterparts of twisted photons -- the vortex electrons \cite{Bliokh0, Bliokh} -- have been generated with the moderately relativistic beams of transmission electron microscopes \cite{Uchida, Verbeeck, McMorran2, Bliokh} and with a non-relativistic beam of a scanning microscope \cite{SEM}. Their OAM can be as high as $1000\hbar$ \cite{l1000, l10002} resulting in a larger magnetic moment compared to that of a plane-wave electron and they can be useful for probing magnetic properties of materials, chirality in crystals, etc. \cite{chir, Bliokh}. These electron packets have also attracted attention outside the microscopy community because of the potential applications in atomic and high-energy physics \cite{Bliokh0, Bliokh, Ll, Ivanov11, I-S, PRA12, Ivanov_PRA_2012, Ivanov12, Surzh, Serbo15, Kaminer2016, Ivanov2016, Ivanov16, JHEP, Sherwin1,  Sherwin2, PRA18, Ivanov20201, Ivanov20202, Ivanov20203, Madan, IvanovMu, Surzhykov21, Peshkov, Silenko23} and, in particular, in hadronic and spin studies -- see the recent review \cite{IvanovRev}. They can even come in handy in accelerator physics \cite{Floettmann, NJP} where their classical counterparts -- \textit{the angular-momentum-dominated beams} -- have been utilized since the beginning of 2000s with the beam vorticity being analogous to the intrinsic OAM of a particle (see \cite{Burov2002, Kim2003, Sun2004, Gr2021} and references therein). 

There are ongoing discussions of the possible experiments with vortex muons, hadrons, ions and atoms, cold neutrons, nuclei, spin waves (magnons), etc. \cite{Ivanov_PRA_2012, Ivanov12, JHEP, PRA18, Ivanov20201, Ivanov20202, Ivanov20203, PRC, Madan, IvanovMu, Floettmann, NJP, Ivanov-atoms, Magnon, Peshkov}. Very recently, in 2021, the non-relativistic twisted atoms and molecules have been generated \cite{Atoms}. Although a large class of quantum and classical waves can carry phase vortices, 
the available diffraction techniques to generate them \cite{Uchida, Verbeeck, McMorran2, Beche, neu} require beams with the large transverse coherence and they are hardly applicable for relativistic beams of accelerators. This circumstance severely limits the development of physics of structured waves to higher energies. 

Indeed, even if the electron beam current is low and there is no space-charge, the Rayleigh length of a relativistic electron packet is usually much larger than the length of the collision region inside a linac, whereas in a storage ring the rms-width $\sqrt{\la \rho^2\ra}$ of a charged-particle packet does not grow but \textit{oscillates} \cite{NJP, FA0, FA1, FA2}, revealing quantum betatron oscillations or broadening of the classical trajectories \cite{ST}. Therefore, the transverse coherence length of a packet \textit{naturally remains small}, and the corresponding quantum effects do not play any role in collider experiments. To bring physics of the non-Gaussian matter waves to the high-energy domain, alternative methods for generating the vortex states of a large class of quantum systems are needed. Several such techniques have been discussed by one of us in \cite{Floettmann, NJP} and it was shown that it is possible to accelerate vortex electrons up to ultrarelativistic energies by modifying the particle sources.
 
In this paper, we further elaborate our ideas from Ref.\cite{Letter} and describe a method to generate vortex states of particles of in principle arbitrary mass, spin, and energy, including hard X-ray and $\gamma$-range photons, muons, protons, ions and nuclei. They can be obtained during emission of photons in accelerators and free-electron lasers, via scattering or annihilation processes with leptons and hadrons, such as 
\be 
e\mu \to e\mu,\, ep \to ep,\, e^-e^+ \to h\bar{h},\, e^-e^+ \to 2 \gamma, 
\ee
scattering of light on relativistic ions, discussed within the Gamma Factory project \cite{Budker, Budker2}, etc. Our key observation is that it is largely \textit{not the process itself} that defines vorticity of a final particle, but rather it is a post-selection protocol due to entanglement between the final particles. Whereas in the classical realm the radiation is twisted if the emitting electron path is helical \cite{Katoh2}, the more general quantum theory developed here predicts that photons \textit{cannot be twisted at all} if the emitting particle 3-momentum $\bp'$ is precisely measured, and that their OAM depends on the way we post-select the electron. Our results complement and specify those of the classical theory, but also make them somewhat less optimistic because the conventional post-selection protocol \cite{BLP, Peskin} results in no vorticity of the emitted photons whatsoever.

We point out a deep analogy between \textit{a generalized measurement} \cite{BarGen, FrankeGen} of a particle momentum ${\bm p}' = \{p'_x, p'_y, p'_z\} = \{p'_{\perp} \cos\phi', p'_{\perp} \sin\phi', p'_z\}$, in which not all the components are measured with a vanishing error, and a standard von Neumann measurement in cylindrical basis \cite{Ivanov_PRA_2012}, in which the azimuthal angle $\phi' = \arctan(p'_y/p'_x)$ \textit{does not belong to the measurable set} of quantum numbers. If this angle is measured with a very large error, this lack of information is enough for the state of the other final particle to become twisted, thanks to entanglement and to the angle-OAM uncertainty relation \cite{Carruthers, Padgett}.  

The crucial advantage of this technique is that neither modifications of the incoming beams are required nor there are limitations due to their small transverse coherence. To generate the vortex state of any particle -- be it a $\gamma$-ray, a proton, or a nucleus in a reaction with two final particles -- it is enough to measure the azimuthal angle of the other particle with a large error or not to measure it at all. This method can be employed for the generation of highly energetic vortex beams at the electron and hadron accelerators, X-ray free-electron lasers and synchrotron radiation facilities with helical undulators such as, for instance, the European XFEL and ESRF, powerful laser facilities such as the Extreme Light Infrastructure, and at the future linear colliders. A system of units with $\hbar = c = 1$ is used.






\section{Evolved state of the final particles}\label{EvSt}

Let us start with emission of a photon by a charged particle, 
\be
e \to e' + \gamma,
\ee 
in the lowest order of the perturbation theory in QED. Generalizations to the scattering and annihilation processes are straightforward. For definiteness, we will take an electron in what follows. It can generate either Cherenkov radiation or transition radiation, synchrotron radiation, undulator radiation in a given electromagnetic field, and so on. A quantum state of the final system as it evolves from the reaction irrespectively of the measurement protocol can be called \textit{pre-selected} or \textit{evolved} because no measurements have been done yet. An initial state $|\text{in}\ra$ of the electron and an evolved state of the final system $|\text{ev}\ra$ are connected via an evolution operator $\hat{S}$ \cite{BLP, Peskin}
\bea
& |\text{ev}\ra = \hat{S}\,|\text{in}\ra,\ \hat{S} = \text{T} \exp\left\{-ie\int d^4x\, \hat{j}_{\mu} \hat{A}^{\mu}\right\}.
\label{Sev}
\eea
In what follows, we restrict ourselves only to the first-order QED processes with $\hat{S}=\hat{S}^{(1)}$ for which the evolved state consists of the electron and the photon, $|\text{ev}\ra = |e', \gamma\ra^{(\text{ev})}$. In this case,
\be
|e', \gamma\ra^{(\text{ev})} = \hat{S}^{(1)}\,|\text{in}\ra = \sum\limits_{f} |f\ra \la f|\hat{S}^{(1)}|\text{in}\ra,
\label{2part}
\ee
where we have expanded the unitary operator over a complete set of two-particle states (with no virtual particles)
\bea
\hat{1} = \sum\limits_{f} |f\ra \la f| = \sum\limits_{f_e}\sum\limits_{f_{\gamma}}|f_e, f_{\gamma}\ra \la f_e, f_{\gamma}|.
\eea
In particular, this can be the plane-wave states of free particles with the definite momenta ${\bm p}, {\bm k}$, the energies $\varepsilon = \sqrt{m_e^2 + {\bm p}^2}, \omega = |{\bm k}|$, and the helicities $\lambda= \pm 1/2, \lambda_{\gamma} = \pm 1$, so that
\be
\hat{1} = \sum\limits_{\lambda_{\gamma}\lambda} \int \frac{d^3p}{(2\pi)^3} \frac{d^3k}{(2\pi)^3} |\bk,\lambda_{\gamma}; \bp,\lambda\ra\la\bp,\lambda;\bk,\lambda_{\gamma}|.
\label{ComplPW}
\ee 
In this case, the average
\be
S_{fi}^{(1)} = \la f_e, f_{\gamma}|\hat{S}^{(1)}|\text{in}\ra = \la\bp,\lambda;\bk,\lambda_{\gamma}|\hat{S}^{(1)}|\text{in}\ra
\label{SPW}
\ee 
is a customary firs-order matrix element with two final plane-wave states. 

Let us first assume that no final particle is detected and discuss how one can describe the entangled evolved state. The field operators in the Heisenberg representation of the photon and of the electron are expanded into series of the creation and annihilation operators of the freely propagating plane-wave states
\bea
&  \displaystyle \hat{\bm A}({\bm r}, t) = \sum\limits_{\lambda_{\gamma}=\pm 1} \int\frac{d^3k}{(2\pi)^3} \left({\bm A}_{{\bm k}\lambda_{\gamma}}({\bm r}, t)\, \hat{c}_{{\bm k}\lambda_{\gamma}} + \text{h.c.}\right),\, {\bm A}_{{\bm k}\lambda_{\gamma}}({\bm r}, t) = \frac{\sqrt{4\pi}}{\sqrt{2\omega}}\, {\bm e}_{{\bm k}\lambda_{\gamma}}\, e^{-i\omega t + i{\bm k}\cdot{\bm r}}, \cr
&  \displaystyle \hat\psi({\bm r}, t) = \sum\limits_{\lambda=\pm 1/2} \int\frac{d^3p}{(2\pi)^3} \psi_{{\bm p}\lambda}({\bm r}, t) \hat{a}_{{\bm p}\lambda},
\, \psi_{{\bm p}\lambda}({\bm r}, t) = \frac{1}{\sqrt{2\varepsilon}}\, u_{{\bm p}\lambda}\, e^{-i\varepsilon t + i{\bm p}\cdot{\bm r}},
\label{FOp}
\eea
where we have omitted the positron part. Note that for Cherenkov radiation in a transparent medium with a refraction index $\tilde n (\omega)$, we have $\omega = |\bm k|/\tilde n$ and the normalization factor must be $1/\sqrt{2\omega}\to 1/\tilde n \sqrt{2\omega}$, see Ref.\cite{Ivanov2016} and Sec.\,4 below. Now, following the standard interpretation \cite{BLP, Peskin, Scully}, one can define the two-particle evolved state in 4-dimensional space-time as follows:
\bea
& \displaystyle \la 0| \hat{\psi}({\bm r}_e, t_e) \hat{{\bm A}}({\bm r}_{\gamma}, t_{\gamma})|e',\gamma\ra^{(\text{ev})} = \cr & \displaystyle = \int\frac{d^3p}{(2\pi)^3}\frac{d^3k}{(2\pi)^3}\,\sum\limits_{\lambda_{\gamma}=\pm 1}\sum\limits_{\lambda=\pm 1/2} \frac{\sqrt{4\pi}}{\sqrt{2\omega}} \frac{1}{\sqrt{2\varepsilon}}\, u_{{\bm p}\lambda}\, {\bm e}_{{\bm k}\lambda_{\gamma}}\, S_{fi}^{(1)}\, e^{-i\varepsilon t_e + i{\bm p}\cdot{\bm r}_e -i\omega t_{\gamma} + i{\bm k}\cdot{\bm r}_{\gamma}},
\eea
with $S_{fi}^{(1)}$ from (\ref{SPW}). This average yields the probability amplitude to detect the photon at the time instant $t_{\gamma}$ in a region of space centered at the point ${\bm r}_{\gamma}$, whereas the electron is detected in the region centered at the point $(t_e,{\bm r}_e)$. So the combination 
\bea
\sum\limits_{\lambda_{\gamma}=\pm 1} \sum\limits_{\lambda=\pm 1/2} u_{{\bm p}\lambda}\, {\bm e}_{{\bm k}\lambda_{\gamma}}\, S_{fi}^{(1)}
\label{APsievk}
\eea
plays a role of the wave function of the 2-particle evolved state in the momentum representation, up to the factor $\sqrt{4\pi}/\sqrt{2\omega} \sqrt{2\varepsilon}$.  

Now let us assume that \textit{only one} of the final particles is detected. Say, if the electron is post-selected to the state $|e'\ra$, the evolved state of the photon alone looks as
\be
|\gamma\ra^{(\text{ev})} = \sum\limits_{f_{\gamma}} |f_{\gamma}\ra \la e'; f_{\gamma}|\hat{S}^{(1)}|\text{in}\ra \equiv \sum\limits_{f_{\gamma}} |f_{\gamma}\ra S_{fi}^{(1)},
\ee
or when the plane-wave basis is used
\be
|\gamma\ra^{(\text{ev})} = \sum\limits_{\lambda_{\gamma}}\int \frac{d^3k}{(2\pi)^3} |{\bm k},\lambda_{\gamma}\ra \la e'; {\bm k},\lambda_{\gamma}|\hat{S}^{(1)}|\text{in}\ra \equiv \sum\limits_{\lambda_{\gamma}}\int \frac{d^3k}{(2\pi)^3} |{\bm k},\lambda_{\gamma}\ra S_{fi}^{(1)}.
\label{gevpw}
\ee
Analogously, if the photon is detected in a state $|\gamma\ra$, the electron evolved state becomes
\be
|e'\ra^{(\text{ev})} = \sum\limits_{f_e} |f_e\ra \la \gamma; f_e|\hat{S}^{(1)}|\text{in}\ra \equiv \sum\limits_{f_e} |f_e\ra S_{fi}^{(1)},
\ee
or
\be
|e'\ra^{(\text{ev})} = \sum\limits_{\lambda}\int \frac{d^3p}{(2\pi)^3} |{\bm p},\lambda\ra S_{fi}^{(1)}.
\ee

One can treat the averages
\bea
&  \displaystyle {\bm A}^{(\text{ev})}({\bm r},t) \equiv \la 0|\hat{\bm A}({\bm r}, t)|\gamma\ra^{(\text{ev})} = \sum\limits_{\lambda_{\gamma}}\int \frac{d^3k}{(2\pi)^3}\, \frac{\sqrt{4\pi}}{\sqrt{2\omega}}\, {\bm e}_{{\bm k}\lambda_{\gamma}}\, \la {\bm k},\lambda_{\gamma}|\gamma\ra^{(\text{ev})}\, e^{-i\omega t + i{\bm k}\cdot{\bm r}},\cr
&  \displaystyle
 \psi^{(\text{ev})}({\bm r},t) \equiv \la 0|\hat\psi({\bm r}, t)|e'\ra^{(\text{ev})} = \sum\limits_{\lambda} \int\frac{d^3p}{(2\pi)^3}\,\frac{1}{\sqrt{2\varepsilon}}\, u_{{\bm p}\lambda} \la {\bm p},\lambda|e'\ra^{(\text{ev})} e^{-i\varepsilon t + i{\bm p}\cdot{\bm r}}
\label{WVxt}
\eea
as the probability amplitudes to detect the corresponding particle in the region of space-time centered at the point $(t, {\bm r})$. In particular, if the photon evolved state is the plane wave, $|\gamma\ra^{(\text{ev})} = |{\bm k}',\lambda'\ra$, then
\bea
\la 0|\hat{\bm A}({\bm r}, t)|{\bm k}',\lambda'\ra = {\bm A}_{{\bm k}'\lambda'_{\gamma}}({\bm r}, t)
\eea
from Eq.(\ref{FOp}) due to the orthogonality condition,
\bea
\la {\bm k},\lambda_{\gamma}|{\bm k}',\lambda_{\gamma}'\ra = \la 0| \hat{c}_{{\bm k}\lambda_{\gamma}}\hat{c}^{\dagger}_{{\bm k}'\lambda'_{\gamma}}|0\ra = (2\pi)^3\delta({\bm k} - {\bm k}')\delta_{\lambda_{\gamma}\lambda'_{\gamma}}.
\label{orth}
\eea
As the plane wave is completely delocalized in time and space, it is generally the function 
\bea
&  \displaystyle \sum\limits_{\lambda_{\gamma}} {\bm e}_{{\bm k}\lambda_{\gamma}}\, \la {\bm k},\lambda_{\gamma}|\gamma\ra^{(\text{ev})} = \sum\limits_{\lambda_{\gamma}} {\bm e}_{{\bm k}\lambda_{\gamma}}\, S_{fi}^{(1)},\ S_{fi}^{(1)} = \la e'; {\bm k},\lambda_{\gamma}|\hat{S}^{(1)}|\text{in}\ra,
\label{Agevk}
\eea
that has a sense of the photon wave function in the momentum representation \cite{BLP}. Analogously, the electron wave function is
\bea
& \displaystyle \sum\limits_{\lambda=\pm 1/2} u_{{\bm p}\lambda} S_{fi}^{(1)},\ S_{fi}^{(1)} = \la \gamma; \bp,\lambda|\hat{S}^{(1)}|\text{in}\ra,
\label{Psievk}
\eea
when the photon detected state $|\gamma\ra$ is given.

Importantly, the 1-particle functions (\ref{WVxt}) can be treated as the probability amplitudes only when the other particle is also detected at a finite point of space-time, and its state represents \textit{a spatially localized} wave packet. Indeed, let us assume the opposite: the incoming electron is a plane wave $|\bp,\lambda\ra$ and the final electron is also detected in the delocalized plane-wave state $|e'\ra = |\bp',\lambda'\ra$ (say, during Cherenkov radiation). Then the photon evolved state becomes
\bea
&  \displaystyle {\bm A}^{(\text{ev})}({\bm r},t) = \sum\limits_{\lambda_{\gamma}}\int \frac{d^3k}{(2\pi)^3}\, \frac{\sqrt{4\pi}}{\sqrt{2\omega}}\, {\bm e}_{{\bm k}\lambda_{\gamma}}\, S_{fi}^{(1)}\, e^{-i\omega t + i{\bm k}\cdot{\bm r}},\cr
&  \displaystyle
S_{fi}^{(1)} = \la \bp',\lambda'; \bk,\lambda_{\gamma}|\hat{S}^{(1)}|\bp,\lambda\ra \propto \delta^{(4)}(p - p'- k). 
\label{Aevrt}
\eea
The integral over $d^3k$ is done with the momenta delta-function, and ${\bm A}^{(\text{ev})}({\bm r},t)$ stays proportional to $\delta(\varepsilon - \varepsilon' - \omega) = (2\pi)^{-1}\int_{t_{\text{in}}=-\infty}^{t_{\text{out}}=+\infty}dt\, \exp\left\{it(\varepsilon - \varepsilon' - \omega)\right\}$,  
due to the infinite limits in the time integral. To circumvent this singular behavior, one can take the final particle (electron or photon) as a localized packet with \textit{a detector function} $f_p(\bp')$,
\bea
&  \displaystyle |e'\ra = \int\frac{d^3p'}{(2\pi)^3} f_p(\bp') |\bp',\lambda'\ra,\cr
&  \displaystyle \la e'|e'\ra = \int\frac{d^3p'}{(2\pi)^3} |f_p(\bp')|^2 = 1,
\label{fp}
\eea
where we have used
\bea
\la {\bm p},\lambda|{\bm p}',\lambda'\ra = (2\pi)^3\delta({\bm p} - {\bm p}')\delta_{\lambda\lambda'}.
\label{orthp}
\eea
Then the evolved states become
\bea
&  \displaystyle {\bm A}^{(\text{ev})}({\bm r},t) = \sum\limits_{\lambda_{\gamma}}\int \frac{d^3k}{(2\pi)^3}\frac{d^3p'}{(2\pi)^3}\,f_p^*(\bp')\, \frac{\sqrt{4\pi}}{\sqrt{2\omega}}\, {\bm e}_{{\bm k}\lambda_{\gamma}}\, S_{fi}^{(1)}\, e^{-i\omega t + i{\bm k}\cdot{\bm r}},\cr
&  \displaystyle \psi^{(\text{ev})}({\bm r},t) = \sum\limits_{\lambda} \int\frac{d^3p}{(2\pi)^3}\frac{d^3k}{(2\pi)^3}\,f_k^*(\bk)\,\frac{1}{\sqrt{2\varepsilon}}\, u_{{\bm p}\lambda} S_{fi}^{(1)} e^{-i\varepsilon t + i{\bm p}\cdot{\bm r}},
\label{WVevg}
\eea
and their momentum counterparts
\bea
&  \displaystyle \int \frac{d^3p'}{(2\pi)^3}\,\sum\limits_{\lambda_{\gamma}} f_p^*(\bp')\, {\bm e}_{{\bm k}\lambda_{\gamma}}\, S_{fi}^{(1)}\quad \text{and}\quad \int \frac{d^3k}{(2\pi)^3}\sum\limits_{\lambda}f_k^*(\bk)\, u_{{\bm p}\lambda} S_{fi}^{(1)},
\label{WVevgp}
\eea
depend on the functions $f_p(\bp'),f_k(\bk)$ describing detector properties of the other particle. As the square-integrable functions $f_p(\bp')$ or $f_k(\bk)$ localize the packets spatially and temporarily, the S-matrix element is effectively ``regularized'', $S_{fi}^{(1)} \to \int \frac{d^3p'}{(2\pi)^3}\, f_p^*(\bp')\, S_{fi}^{(1)}$, so that the evolution now occurs \textit{in a finite region of space and during a finite time}. The functions (\ref{WVevgp}) keep their sense in the limit $f_p(\bp') \to (2\pi)^3 \delta (\bp' - \bar{\bp}'), f_k(\bk) \to (2\pi)^3 \delta (\bk - \bar{\bk})$ when they reduce to Eq.(\ref{Agevk}) and (\ref{Psievk}), respectively. In this paper, we do not aim at a full description of the detection process in space-time and, therefore, we will mostly study the states in the momentum representation for which $f_p(\bp'),f_k(\bk)$ may not be square-integrable.

Clearly, the above definitions can easily be generalized for scattering and annihilation processes, both elastic and inelastic, including reactions with hadrons, ions, or nuclei in the final state. In the latter case, Eq.(\ref{Psievk}) can define an evolved wave function of a hadron when all other particles are detected.

We emphasize the difference between the above quantum picture and the classical theory: in the latter the electron path is given, the particle experiences no recoil, the coherent properties and the phase of the radiation field are solely defined by the trajectory. In particular, when the electron path is spiral the radiation field is always twisted \cite{Katoh2}. The quantum picture \textit{embraces the classical problem} as a special case (the correspondence principle), but it is inevitably more complex: the evolved state of the emitted photon always depends on how the emitting particle (electron) is post-selected. As we show below, if the electron 3-momentum is measured via the standard projective measurement approach \cite{Peskin}, the final photon turns out to be \textit{not twisted at all}, even if the electron packet centroid moves along a helix. Thus, the twisted particles may not be \textit{that abundant} in Nature as it seems to follow from the classical theory.

\section{Measurement scenarios} 
\subsection{Projective measurements in plane-wave basis}

Let the initial electron be described as a plane-wave state propagating along the $z$ axis with the momentum $\bp = \{0,0,|\bp|\}$ (say, for Cherenkov radiation). Following the conventional textbook approach \cite{BLP, Peskin}, if it is post-selected to a plane-wave state with the momentum $\bp' = \{\bp'_{\perp}, p'_z\}$ and the helicity $\lambda'$, the matrix element $S_{fi}^{(\text{pw})}$ and the photon's evolved wave function are both proportional to the momentum conservation delta-function
\bea
& S_{fi}^{(\text{pw})} \propto \delta^{(3)} (\bp - \bp' - \bk) \propto \delta^{(2)} (\bp'_{\perp} + \bk_{\perp}),\cr 
& {\bm A}^{(\text{ev})}(\bk, \omega) \propto \sum\limits_{\lambda_{\gamma} = \pm 1} {\bm e}\, \delta (\bp'_{\perp} + \bk_{\perp}).
\label{evA}
\eea
As soon as the azimuthal angle $\phi'$ of the electron momentum is precisely measured, the photon angle $\phi_k$ is also set to a definite value
\be
\phi_k = \phi' \pm \pi. 
\ee
Thereby, the photon is projected to a plane-wave state \textit{without} an intrinsic OAM projection on the $z$ axis.
One can qualitatively understand this result by recalling the uncertainty relation from Ref.\cite{Carruthers, Padgett}, which can be put as follows: \textit{a definite azimuthal angle of a plane wave implies a vanishing mean OAM $z$-projection $\la\hat{L}_z\ra = 0$ but an infinitely wide OAM spectrum $\la\hat{L}^2_z\ra = \infty$} (see appendix \ref{AppA} for more detail).	

Now let us take a general reaction with the plane-wave states, the matrix element of which is proportional to the energy-momentum conservation delta-function \cite{BLP, Peskin},
\bea
& S_{fi}^{(\text{pw})} \propto \delta^{(4)} \left(\sum p_{\text{in}} - \sum p_{\text{out}}\right).
\label{Sgen}
\eea
If we substitute this to Eq.(\ref{Agevk}) or (\ref{Psievk}), we immediately see that the evolved wave function of a photon or of a fermion has a well-defined 4-momentum, which automatically means that $\la\hat{L}_z\ra = 0$.

The projective measurements in the plane-wave basis are most frequently used for describing particle scattering, annihilation, or photon emission. As we have said, in quantum theory it is not the particle trajectory but the post-selection protocol that defines vorticity of the final photon. Therefore, when it comes to the photon phase, direct comparison of the results in the plane-wave basis with the classical theory \textit{has a limited sense}. Indeed, the classical electron does not experience recoil, so its angle $\phi'$ \textit{is not defined}, which is more similar to the generalized-measurement scheme \textit{(iii)} below than to the above standard approach.


\subsection{Cylindrical basis and generalized measurements}\label{Sch}

The post-selection to a plane wave represents a projective measurement when all components of the momentum $\bp'$ are measured with vanishing errors. During a generalized measurement, on the contrary, at least some of the components are measured with a \textit{non-vanishing} error (see, e.g., \cite{BarGen, FrankeGen}). If the error is maximal, we do not effectively measure this component at all. Put simply, a generalized measurement is the post-selection to a wave packet, a coherent superposition of plane waves,
\bea
|e'_{\text{det}}\ra^{(\text{g})} = \int\fr{d^3p'}{(2\pi)^3}\, f_p(\bp')\, |\bp',\lambda'\ra,
\label{wp}
\eea
where the detector function $f_p(\bp')$ is normalized according to Eq.(\ref{fp}).
The set of operators 
\bea
|{e'_{\text{det}}\ra}^{(\text{g})}\la {e'_{\text{det}}}|^{(\text{g})}
\label{POVM}
\eea 
is complete, although \textit{not necessarily orthogonal}, and it forms a positive operator-valued measure \cite{BarGen}.
There exists a close relative of the generalized measurement, called \textit{the weak measurement} \cite{weakPRL, weak1, weak2, weak3},
during which an observable quantity is also measured with a finite error, but the measured \textit{weak values} do not necessarily belong to the spectrum of the corresponding operator and they can even be complex. The operators (\ref{POVM}), needed for our purposes, are hermitian and positively-defined, so the measured values of the azimuthal angle still belong to the interval $[0, 2\pi]$, which is why we adhere to the term ``generalized measurement'' in what follows.

\begin{figure*}[!htb]
   \centering
	\includegraphics[width=15cm]{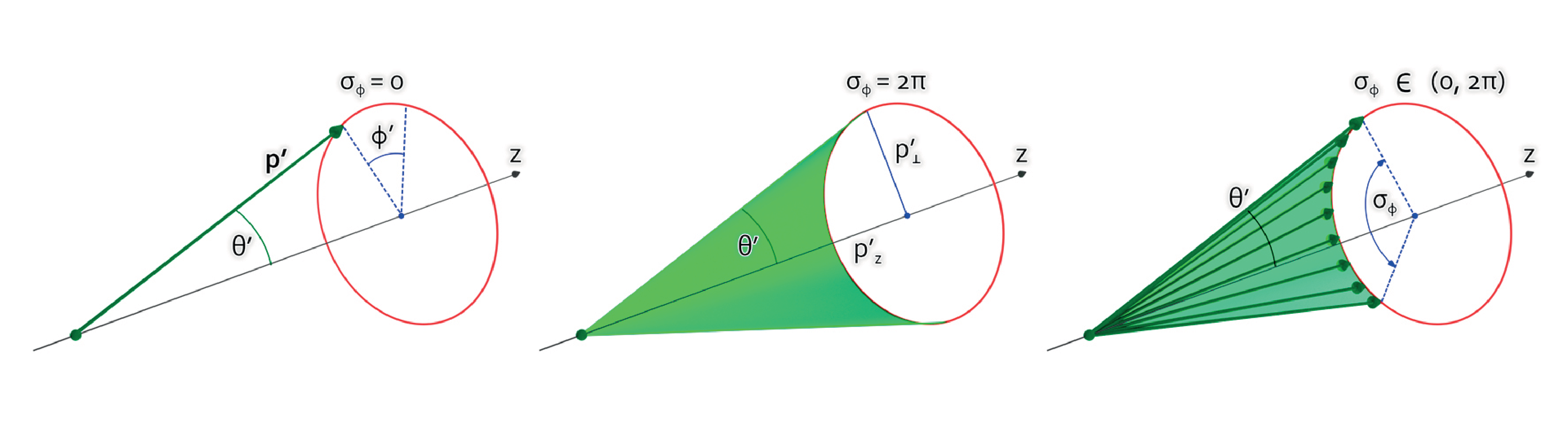} 
	\caption{The possible measurement strategies. Left: the conventional plane-wave approach \textit{(i)} (projective measurement) with the definite 3-momentum $\bp'$, middle: the projective measurement in the cylindrical basis \textit{(ii)} with an undefined azimuthal angle $\phi'$ of the momentum, right: the generalized-measurement scheme \textit{(iii)} with a finite uncertainty $\sigma_{\phi}$ of the azimuthal angle. In all three scenarios, the states are stationary with the definite energy, $\varepsilon' = \sqrt{(p'_{\perp})^2 + (p'_z)^2 + m_e^2}$.
\label{Fig1}}
\end{figure*}

The detector function can be of the Gaussian form, 
\be
f_p(\bp') \propto \prod\limits_{i}\exp\left\{-(p'_i - \la p_i\ra)^2/(2\sigma_i)^2\right\}.
\ee
where $i=x,y,z$ or $i=\rho,\phi,z$. A projective measurement implies that $\sigma_x, \sigma_y, \sigma_z \to 0$. Alternatively, one can use the cylindrical coordinates $p'_{\perp}, \phi' = \arctan(p'_y/p'_x), p'_z$ with the corresponding uncertainties $\sigma_{\perp}, \sigma_{\phi}, \sigma_z$. For a generalized measurement, at least one of these uncertainties is not vanishing. 

To provide completeness of the set (\ref{POVM}), one generally needs to measure both the non-commutative observables, the simplest example being the coordinates and momenta.
The corresponding normalized wave packet can be written as (we omit the helicity sign for now)
\bea
& \displaystyle |e'_{\text{det}}\ra^{(\text{g})} = \left|\la{\bm r}\ra, \la{\bm p}\ra\right\ra = \cr
& \displaystyle = \int\frac{d^3p'}{(2\pi)^3}\, \prod\limits_{i}\frac{(2\pi)^{3/2}}{(2\pi\sigma_i^2)^{1/4}}\,e^{-\frac{\left(p'_i - \la p_i\ra\right)^2}{(2\sigma_i)^2} - i\la{\bm r}\ra\cdot {\bm p}'} |{\bm p'}\ra,\cr
& \displaystyle \left\la\la{\bm r}\ra, \la{\bm p}\ra|\la{\bm r}\ra, \la{\bm p}\ra\right\ra = 1,\
\left\la\la{\bm r}\ra, \la{\bm p}\ra|\hat{{\bm p}}|\la{\bm r}\ra, \la{\bm p}\ra\right\ra = \la{\bm p}\ra,\
\left\la\la{\bm r}\ra, \la{\bm p}\ra|\hat{{\bm x}}|\la{\bm r}\ra, \la{\bm p}\ra\right\ra = \la{\bm r}\ra, 
\label{wpxp}
\eea
and this set is complete
\bea
& \displaystyle \int d^3\la{\bm r}\ra \frac{d^3\la{\bm p}\ra}{(2\pi)^3}\, |\la{\bm r}\ra, \la{\bm p}\ra\ra \la\la{\bm r}\ra, \la{\bm p}\ra| = \int\frac{d^3p'}{(2\pi)^3}\, |{\bm p}'\ra\la {\bm p}'| = \hat{1}.
\label{compl}
\eea
These packets represent the so-called \textit{coherent states} of a freely propagating massive particle \cite{BagrovUFN}.

To illustrate these ideas, let us distinguish the following three scenarios (see Fig.\ref{Fig1}): 

\textit{(i)} We post-select the particles to the plane-wave states and repeat the measurements many times with an ensemble of electrons, each time fixing the detector at a different angle $\phi'$. The emission rate or a cross section are proportional to
\bea
\int\limits_0^{2\pi}\fr{d\phi'}{2\pi}\,|S_{fi}^{\rm(pw)}|^2,
\label{S1}
\eea
which represents \textit{an incoherent} averaging over the azimuthal angle in the projective-measurement scheme. This is the standard textbook approach \cite{BLP, Peskin}, in which no phases contribute to the observable quantities.

\textit{(ii)} Another example of a projective measurement is post-selection to a cylindrical wave (a Bessel state) \cite{Ivanov_PRA_2012} with the definite $p'_{\perp}, p'_z$, the $z$-projection of \textit{the total angular momentum} (TAM) $m'$, and the helicity $\lambda'$, but undefined angle $\phi'$
\bea
& |e'_{\text{det}}\ra = |p'_{\perp}, p'_z, m', \lambda'\ra = \cr
& = \int\limits_0^{2\pi}\fr{d\phi'}{2\pi}\,i^{-(m'-\lambda')} e^{i(m' - \lambda')\phi'}\,|\bp', \lambda'\ra.
\label{cylexp}
\eea
The TAM z-projection operator 
\bea
& \hat{j}_z = \hat{L}_z + \hat{s}_z,\cr
& \hat{j}_z |p'_{\perp}, p'_z, m', \lambda'\ra = m' |p'_{\perp}, p'_z, m', \lambda'\ra,
\eea 
is a sum of an orbital part $\hat{L}_z$ and a spin part $\hat{s}_z$.
The former is 
\be
\hat{L}_z = [\hat{\br}\times \hat{\bp}]_z = -i\frac{\partial}{\partial \phi},
\ee
while the latter depends on the particle spin. Here, $\phi$ refers to the azimuthal angle either of $\br$ or of $\bp$, depending on the representation.  

The corresponding amplitude
\bea
\int\limits_0^{2\pi}\fr{d\phi'}{2\pi}\,i^{m'-\lambda'} e^{-i (m'-\lambda') \phi'}\,S_{fi}^{\rm (pw)}
\label{S2}
\eea
represents \textit{a coherent} averaging over the azimuthal angle, whereas the detector is able to measure the TAM with a vanishing error $\sigma_{m} \to 0$. As $\phi$ and $m$ represent \textit{the conjugate variables} \cite{Carruthers}, in this way we also obtain complete information about the electron, but in the cylindrical basis.

\textit{(iii)} Consider now an electron emitting a photon when we measure the angle $\phi'$ of the final electron momentum with a finite error (see the right panel in Fig.\ref{Fig1}) 
\be
\sigma_{\phi} \in (0,2\pi).
\ee
In this scheme we post-select to the packet (\ref{wp}) with the finite uncertainties $\sigma_{\perp}, \sigma_{\phi}, \sigma_z$,
and the non-vanishing uncertainty $\sigma_{\phi}$ means that we do not know exactly at which azimuthal angle the electron goes,
whereas the maximal value, $\sigma_{\phi} \to 2\pi$, corresponds to the special case when the angle is not known at all or -- simply -- is not measured.
The uncertainty relation \cite{Carruthers} says that the larger the error of the angle is, the smaller is the corresponding OAM error 
in the subsequent measurement. Thus, the generalized measurement scheme employs the detector function $f(\phi, \sigma_{\phi})$ from Eq.(\ref{WVevgp}), and the evolved states are
\bea
{\bm A}^{(\text{ev})}_{(\text{g})}(\bk, \omega) = \sum\limits_{\lambda_{\gamma} = \pm 1} {\bm e}\, \int\limits_0^{2\pi}\fr{d\phi'}{2\pi}\,f(\phi', \sigma_{\phi})\, S_{fi}^{(\text{pw})},\cr
\psi^{(\text{ev})}_{(\text{g})}(\bp', \varepsilon') = \sum\limits_{\lambda' = \pm 1/2}\, u'\, \int\limits_0^{2\pi}\fr{d\phi'}{2\pi}\,f(\phi', \sigma_{\phi})\, S_{fi}^{(\text{pw})},
\label{EvWVgen}
\eea
where $f(\phi', 0) = \delta (\phi' - \tilde{\phi}')$ and $f(\phi', 2\pi) = 1$. When $|\sigma_{\phi} - 2\pi| \ll 2\pi$, the function $f$ is smooth, 
and the states represent the Bessel-like wave packets with a mean TAM $\la \hat{j}_z\ra$ and a \textit{finite} TAM dispersion, $\la \hat{j}_z^2\ra - \la \hat{j}_z\ra^2 \ne 0$. 

\begin{figure}[t]
	\center
	\includegraphics[width=0.5\linewidth]{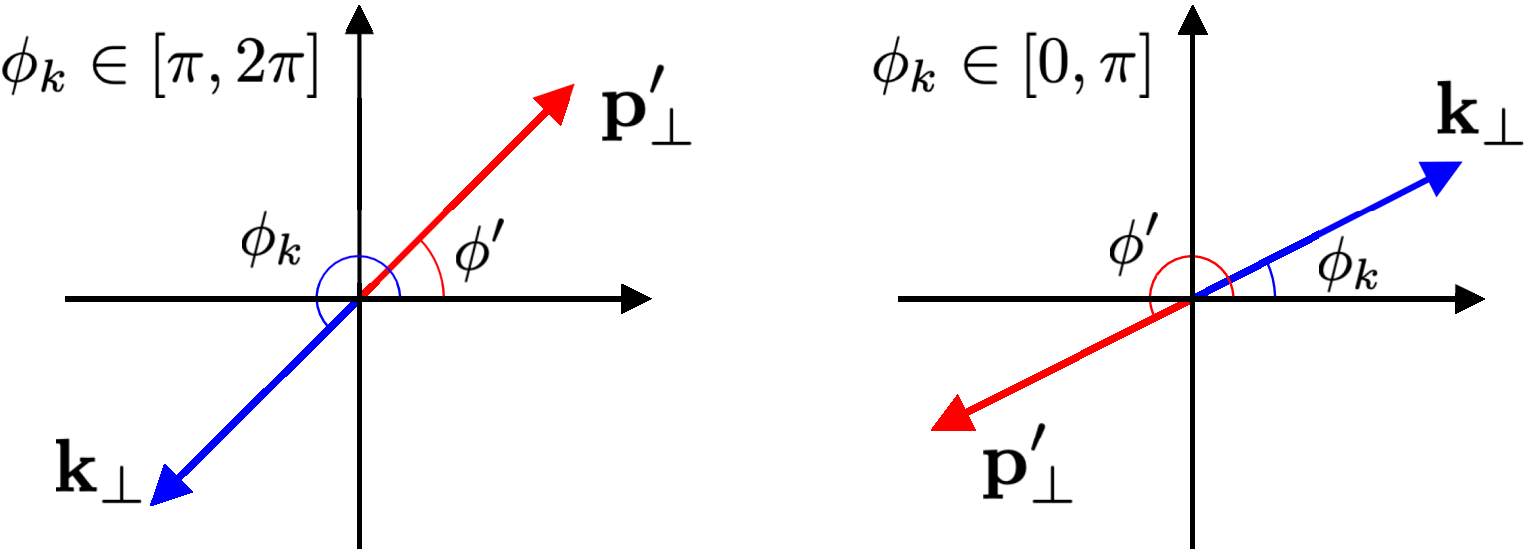}
	\caption{Two possible directions of the vectors $\mathbf{k}_\perp$ and $\mathbf{p}'_\perp$ according to Eq.\eqref{eq:deltaexp}.
\label{Fig2}}
\end{figure}

In what follows, we discuss the scheme with $\sigma_{\phi} \to 2\pi$ for simplicity for which the information about the electron momentum is \textit{incomplete}, but the energy is well-defined, $\varepsilon' = \sqrt{(p'_{\perp})^2 + (p'_z)^2 + m_e^2}$. The corresponding amplitude is also obtained via the coherent averaging ($f=1$), 
\bea
\int\limits_0^{2\pi}\fr{d\phi'}{2\pi}\,S_{fi}^{\rm(pw)}.
\label{S3}
\eea
Formally, this expression coincides with (\ref{S2}) at $m'-\lambda'=0$, but its physical meaning is different. In the scheme \textit{(ii)}, we do measure the TAM projection and the helicity, and we can easily obtain $\la \hat{L}_z\ra = m'-\lambda'=0$, but during the generalized measurement we do not effectively measure the TAM at all, which implies projection to the state 
\be
|e'_{\text{det}}\ra^{\text{(g)}} = 
\int\limits_0^{2\pi}\fr{d\phi'}{2\pi}\,|\bp', \lambda'\ra,
\label{cyl0}
\ee
where each plane wave enters with the same phase. The corresponding function $f_p(\bp')$ from Eq.(\ref{wp})
\bea
f_p(\bp') = (2\pi)^2 \delta \left(p'_z - \la p\ra_z\right) \frac{1}{p'_{\perp}} \delta \left(p'_{\perp} - \la p\ra_{\perp}\right)
\label{psipgen}
\eea
does not depend on the angle $\phi'$ and it must be understood as a limiting case of a wave packet.

\begin{figure}[t]
	\hspace*{-0.2cm}
	\center
	\includegraphics[width=0.65\linewidth]{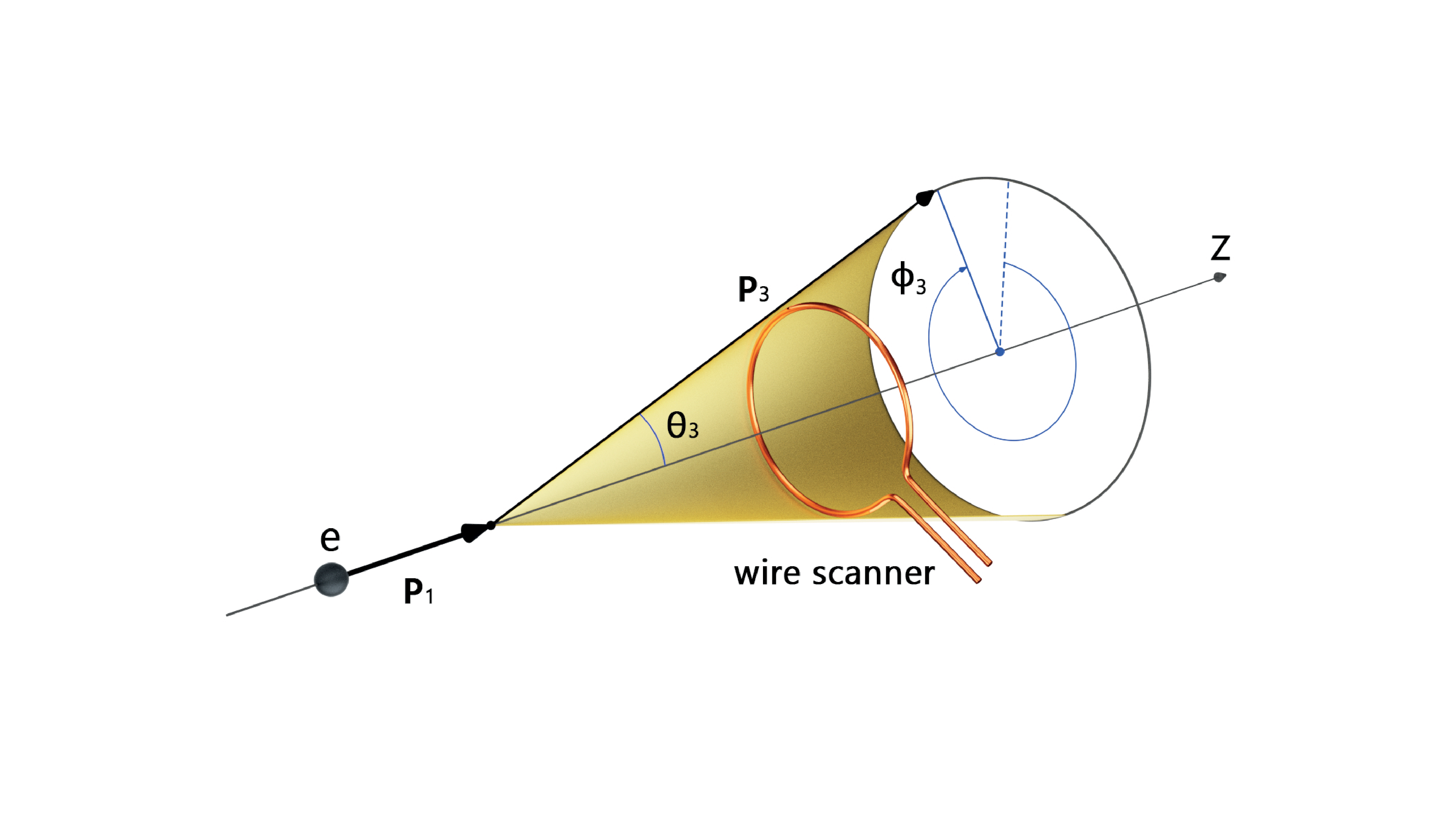}
	\caption{A ring-shaped wire scanner detects electrons scattered at a certain polar angle $\theta_3$, but does not measure the azimuthal angle $\phi_3$.
\label{Fig3}}
\end{figure}

Working in cylindrical coordinates, we use the following representation (see Fig.\ref{Fig2}):
\bea
\label{eq:deltaexp}
& \delta (\bp'_{\perp} + \bk_{\perp}) = \delta (p'_x + k_x)\delta (p'_y + k_y) = \cr & \fr{1}{p'_{\perp}}\,\delta (p'_{\perp} - k_{\perp}) \left(\delta(\phi' - (\phi_k -\pi))\Big|_{\phi_k \in [\pi,2\pi]} + \delta(\phi' - (\phi_k +\pi))\Big|_{\phi_k \in [0,\pi]}\right). 
\label{deltatr}
\eea
Note that only one of these configurations contributes to the evolved state, and there is no interference between them (cf. Ref.\cite{Ivanov16, Ivanov-atoms, Ivanov20201}). Thus, in the simplest scheme with $\sigma_{\phi} \to 2\pi$ the evolved wave functions of the photon and of the fermion become
\bea
{\bm A}^{(\text{ev})}_{(\text{g})}(\bk, \omega) = \sum\limits_{\lambda_{\gamma} = \pm 1} {\bm e}\, \int\limits_0^{2\pi}\fr{d\phi'}{2\pi}\, S_{fi}^{(\text{pw})} \propto \cr
\propto \int\fr{d\phi'}{2\pi}\,\delta (\bp'_{\perp} + \bk_{\perp}) \propto \fr{1}{p'_{\perp}}\,\delta (p'_{\perp} - k_{\perp}),\cr
\psi^{(\text{ev})}_{(\text{g})}(\bp', \varepsilon') = \sum\limits_{\lambda' = \pm 1/2}\, u'\, \int\limits_0^{2\pi}\fr{d\phi'}{2\pi}\, S_{fi}^{(\text{pw})}.
\label{EvWV}
\eea
where for the fermion the integration is performed over the azimuthal angle of the particle measured in the generalized scheme.
The proportionality to $\delta (p'_{\perp} - k_{\perp})$ instead of $\delta (\bp'_{\perp} - \bk_{\perp})$ is a hallmark of the Bessel state 
with the vanishing TAM-uncertainty. In Fig.\ref{Fig3} we show one of the simplest examples how one can detect a charged particle scattered at a certain polar angle, but deliberately do not get any information about its azimuthal angle. 

Finally, note that this generalized-measurement scenario can, somewhat surprisingly, guarantee the validity of \textit{the Bohr correspondence principle} in the quasi-classical regime when the emitting particle experiences very small recoil and the scattering angle is vanishing, $\theta' \ll 1$. For instance, in a number of processes, including Cherenkov radiation, synchrotron radiation, and the Compton effect, this angle can be of the order of 
\be
\theta' \sim 10^{-6} - 10^{-5}\, \text{rad}
\ee
for electrons and realistic parameters. Therefore, precise measurements of the angle $\phi'$ can technically be \textit{very challenging} at these small polar angles. This is the reason why predictions of this genuinely quantum generalized-measurement scheme provide coincidence with the classical theory.

\section{Example 1: Cherenkov radiation by a plane-wave or twisted electron}

\begin{figure*}[t]
	\hspace*{0.2cm}
	\center
\includegraphics[width=0.8\linewidth]{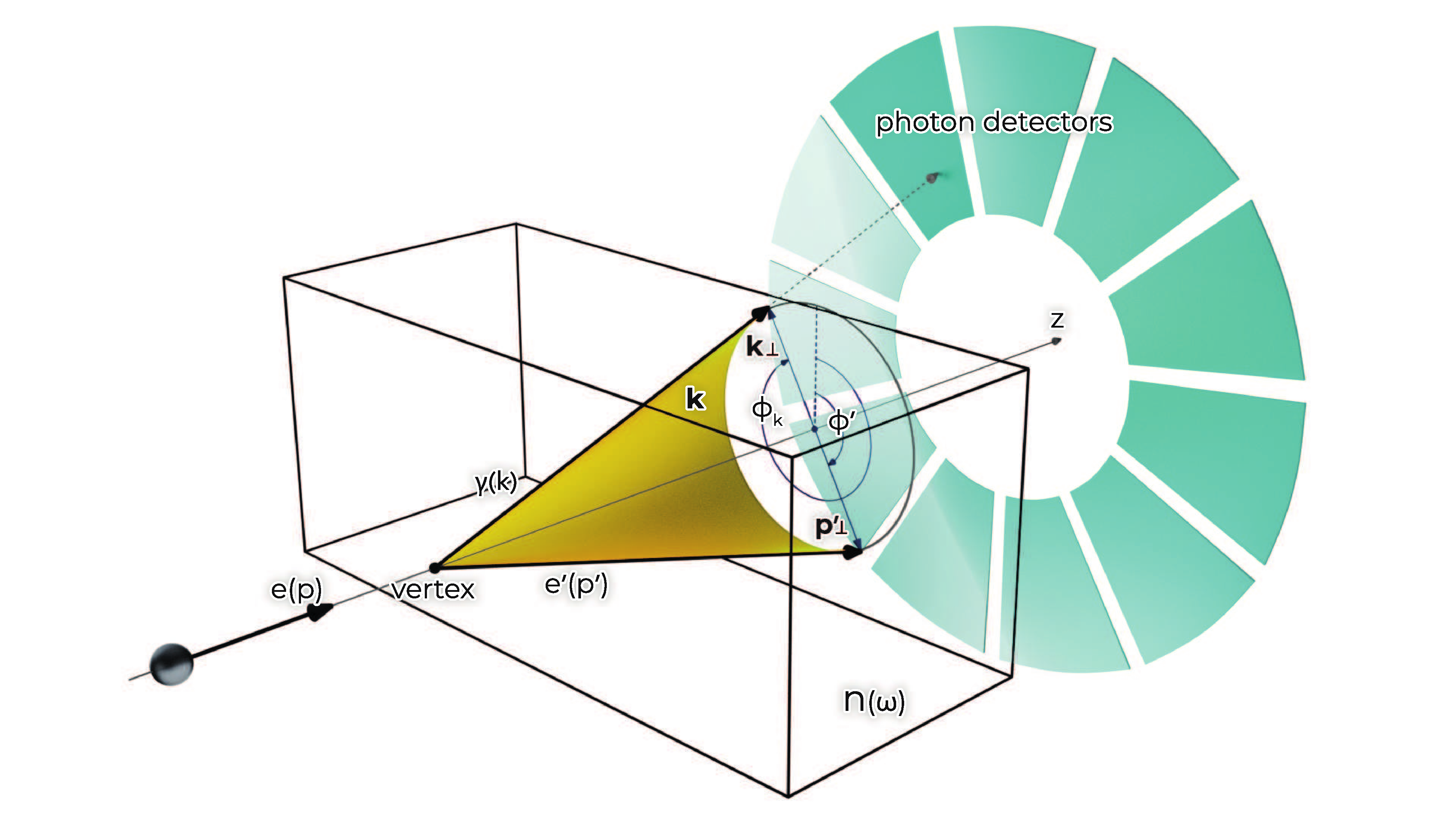}
\caption{An azimuthal ``which-way'' experiment with Cherenkov radiation: during a projective measurement of the electron momentum $\bp'$ we learn the photon angle $\phi_k$, which is why the photon becomes a plane wave. However, if the electron angle $\phi'$ is measured with a large error or is not measured at all, the angle $\phi_k$ stays undefined, and the photon evolved state automatically becomes twisted. In practice, the electron scattering angle is usually very small, $\theta' \ll 1$, so precise measurements of the angle $\phi'$ can be very challenging. In this quasi-classical regime the photons naturally become twisted being coherently emitted to all the azimuthal angles on the Cherenkov cone.}
\label{Fig4}
\end{figure*}

Consider Cherenkov emission, 
\be
e(p) \to e'(p') + \gamma(k), 
\ee
in a transparent medium with weak frequency dispersion and a refractive index $\tilde{n}(\omega)$; see Fig.\ref{Fig4} \cite{Letter}. The incoming electron is described as a plane wave and the matrix element is\footnote{The transition current in the TAM basis and the explicit expressions for the 4-spinors are given in the Appendix \ref{Phases}.}
\bea
& S_{fi}^{\rm(pw)} = -ie N (2\pi)^4 \delta^{(4)}(p-p'-k)\, \bar{u}'\gamma^{\mu}u\, e^*_{\mu},
\eea
where $N$ is a normalization constant (see Ref.\cite{Ivanov2016}) and
\bea
& p = \{\varepsilon, 0, 0,|\bm p|\},\ p' = \{\varepsilon', p'_{\perp}\cos\phi', p'_{\perp}\sin\phi', p'_z\},\cr 
& p'_{\perp} = |\bm p'| \sin \theta',\ k = \{\omega, k_{\perp} \cos\phi_k, k_{\perp} \sin\phi_k, k_z\},\cr 
& |\bk| = \sqrt{k_{\perp}^2 + k_z^2} = \omega\, \tilde{n}(\omega),\, k_{\perp} = |\bk| \sin \theta_k.
\eea
The phases are chosen so that the electron bispinors are eigenfunctions of the TAM z-projection operator with the eigenvalues $\lambda = \pm 1/2, \lambda'= \pm 1/2$.


If the electron is detected in the above generalized-measurement scheme, we do not know the photon momentum azimuthal angle. Then, the photon evolved state becomes
\bea
& {\bm A}^{(\text{ev})}(\bk, \omega) = \int\limits_0^{2\pi}\fr{d\phi'}{2\pi} \sum\limits_{\lambda_{\gamma} = \pm 1} {\bm e}\, S_{fi}^{\rm(pw)} = - ieN {\bm n}\times\left[{\bm n}\times \int\limits_0^{2\pi}\fr{d\phi'}{2\pi} (2\pi)^4 \delta^{(4)}(p-p'-k)\, \bar{u}'{\bm \gamma}u\right],
\label{Phcyl}
\eea
where ${\bm n} = \bk/|\bk|$, we employ the Coulomb gauge with $e^{\mu} = \{0,{\bm e}\}$, and the summation over the  photon helicities is done with the following formula 
\be
\sum\limits_{\lambda_{\gamma}=\pm 1} e_i e^*_j = \delta_{ij} - n_i n_j.
\label{sumhel}
\ee 
Substituting Eq.(\ref{deltatr}) to Eq.(\ref{Phcyl}), we arrive at (see technical details and notations in the Appendix \ref{Phases})
\bea
& {\bm A}^{(\text{ev})} = (-1)^{\lambda-\lambda'} i e N (2\pi)^3 \delta(\varepsilon - \varepsilon' -\omega) 
\delta (|\bm p| - p_z' - k_z)\, \fr{1}{p'_{\perp}}\, \delta (p'_{\perp} - k_{\perp})\cr
& \times\,\left(\sqrt{\varepsilon' + m_e}\sqrt{\varepsilon - m_e} + 2\lambda\,2\lambda'\sqrt{\varepsilon + m_e}\sqrt{\varepsilon' - m_e}\right)
\left[\bm F - \bm n (\bm n \bm F)\right],
\label{PolA}
\eea
where
\bea
& \bm F =d_{\lambda\lambda'}^{(1/2)}(\theta')\, {\bm \chi}_{0}\,  
e^{i(\lambda - \lambda')\phi_k} + \sqrt{2}\,d_{-\lambda\lambda'}^{(1/2)}(\theta')\,{\bm \chi}_{2\lambda} 
e^{-i(\lambda + \lambda')\phi_k}.
\eea
Thus, the two terms with ${\bm \chi_0}\,e^{i(\lambda - \lambda')\phi_k}$ and ${\bm \chi_{2\lambda}}\,e^{-i(\lambda + \lambda')\phi_k}$ in 
$\bm F$ are eigenvectors of $\hat{s}_z$ operator with the eigenvalues $0$ and $2\lambda$, respectively, and the eigenfunctions of the OAM projection operator $\hat L_z= -i \partial/\partial \phi_k$ with the eigenvalues $\lambda-\lambda'$ and $-\lambda-\lambda'$, respectively. Therefore, the vector $\bm F$ itself is an eigenvector of the photon TAM projection operator $\hat{j}^{(\gamma)}_z = \hat{s}_z + \hat{L}_z $ with the eigenvalue $\lambda- \lambda'$.

Similarly, using the representation
\be 
{\bm n} = \sum\limits_{\sigma=0,\pm 1}d^{(1)}_{\sigma 0}(\theta_k){\bm\chi}_{\sigma} e^{-i\sigma\phi_k}
\ee
it is not difficult to prove that the vector $\bm n (\bm n \bm F)$ and, therefore, the photon state $\bm A^{(\text{ev})}$ itself are eigenvectors of $\hat j^{(\gamma)}_z$ operator with a definite value of the photon TAM projection:
\bea
& \hat j_z^{(\gamma)}\bm A^{(\text{ev})} =(\lambda-\lambda') \bm A^{(\text{ev})},
\eea
where
\bea
& \lambda - \lambda' = 
\begin{cases}
    0,\ \text{if}\  \lambda' = \lambda\ \text{(no electron TAM flip)},\\
    2\lambda = \pm 1,\ \lambda' = -\lambda\ \text{(an electron TAM flip)}.
  \end{cases}
\label{TAMcl1}
\eea
So, the maximum TAM value is $|j_z^{(\gamma)}| = 1$. The transverse momentum $k_{\perp}$ of this \textit{Bessel beam} equals to that of the final electron and, therefore, it can be written via the electron scattering angle $\theta'$,
\be
k_{\perp} = p'_{\perp} = |\bm p'| \sin\theta'.
\ee
As this angle is usually negligibly small for Cherenkov radiation, $\theta' \ll 1$, it is very challenging to precisely measure the azimuthal angle $\phi'$. In this sense, Cherenkov radiation with no post-selection of $\phi'$ can be viewed as a natural source of twisted photons, coherently emitted to all the azimuthal angles on the Cherenkov cone. We emphasize that even the state with the vanishing eigenvalue of $\hat j_z^{(\gamma)}$ operator in Eq.(\ref{PolA}) is \textit{not} a plane wave but a cylindrical one with the \textit{spin-orbit interaction} (SOI). 

Clearly, Eq.(\ref{TAMcl1}) simply represents the conservation law of the TAM $z$-projection $\hat{j}_z=\hat{j}'_z+\hat{j}^{(\gamma)}_z$. Now recall that this conservation law takes place not only for the pure states with $\la\hat j_z\ra = \pm 1/2$, but also for the mixed states with $\la\hat{j}_z\ra \in [-1/2,1/2],\, \la\hat{j}'_z\ra \in [-1/2,1/2]$. In the latter case, instead of Eq.(\ref{TAMcl1}) we have
\bea
\la\hat{j}_z\ra^{(\gamma)} = \la\hat{j}_z\ra - \la\hat{j}'_z\ra,
\label{TAMcl}
\eea
where for the mixed electron states the photon TAM lies in the interval $\la\hat{j}_z\ra^{(\gamma)} \in [-1,1]$. In particular, when the initial electron is unpolarized, $\la \hat{j}_z\ra = 0$, and the final electron TAM is not measured, $\la \hat{j}'_z\ra = 0$, the photon TAM is also vanishing, although it is not a plane wave.

Finally, when the initial electron is in a pure \textit{twisted} state \cite{Bliokh} with the TAM $z$-projection 
\be
m = \pm 1/2, \pm 3/2, ...,
\ee 
its bispinor $u \equiv u_{p \lambda}$ with $\bm p = \{p_\perp\cos\phi, p_\perp\sin\phi,p_z\}$ transforms as (see Eq.~\eqref{cylexp}) 
\be
u_{p\lambda} \to u_{p_\perp p_z m \lambda}=
i^{-(m-\lambda)}\int\limits_0^{2\pi} \fr{d\phi}{2\pi} 
\,e^{i(m-\lambda)\phi}\,u_{p \lambda}.
 \label{u-twisted}
\ee
The further integration over $\phi$ and $\phi'$ can be performed analytically. As a result, the photon TAM transforms as $\lambda- \lambda' \to m-\lambda'$: 
\be
\hat j_z^{(\gamma)}\bm A^{(\text{ev})} =(m-\lambda') \bm A^{(\text{ev})}.
\label{TAMcm}
\ee
Thus, \textit{vorticity of the incoming electron can be transferred to the photon} within the generalized measurement scheme, and the twisted photons with the TAM $|j_z^{(\gamma)}| > 1$ can be emitted. Clearly, the similar post-selection protocol can also be applied for other processes, including synchrotron radiation, transition radiation, diffraction radiation, Smith-Purcell radiation, and so on.

\section{Example 2: non-linear Compton scattering \\ and undulator radiation} \label{Compt}
\subsection{Post-selection to the plane-wave state}

In a circularly polarized laser wave with the following potential \cite{BLP, R}
\bea
& A^{\mu} = a_1^{\mu} \cos(kx) + a_2^{\mu} \sin(kx),\cr
& A^2 = a_1^2 = a_2^2 = -a^2 < 0,\cr
& (a_1 a_2) = (ka_1) = (ka_2) = 0,\ kx = \omega t - \bk \cdot \br
\eea
an electron is conventionally described with a Volkov state \cite{BLP, R, Piazza, Fedotov}
\bea
&\psi_{p\lambda}(\br,t) = N_e \left(1 + \fr{e}{2(pk)} (\gamma k)(\gamma A)\right)\, u\, \exp\left\{-ipx - \fr{ie}{(pk)} \int\limits^{kx}d\varphi \left((pA) - \fr{e}{2}A^2\right)\right\},
\label{Volkov}
\eea
which is exact solution of the Dirac equation and where $\bar{u}u = 2m_e$, $N_e$ is a normalization constant, and the second term in the pre-exponential factor is due to the spin,
\bea
& (\gamma k)(\gamma A) = \fr{1}{2}\,\int\limits^{kx}d\varphi\,F^{\mu\nu}\sigma_{\mu\nu},\cr
& \sigma_{\mu\nu} = \fr{1}{2}\left(\gamma^{\mu}\gamma^{\nu} - \gamma^{\nu}\gamma^{\mu}\right),\ 
F^{\mu\nu} = \partial^{\mu}A^{\nu} - \partial^{\nu}A^{\mu}.
\label{spin}
\eea
Note that the Volkov state (\ref{Volkov}) transforms into a plane wave when the laser field is off, $A \to 0$.
The case when the incoming electron is twisted and is in a Bessel-Volkov state \cite{PRA12} is studied hereafter.
The final photon wave function is also a plane wave,
\be
{\mathcal A}^{\mu} = N_{\gamma}\, e'^{\mu}\, e^{-i\omega't + i\bk'\cdot\br},\ k'_{\mu}e'^{\mu} = 0.
\ee
The corresponding matrix element is
\bea
S_{fi}^{(\text{pw})} = -ie\int d^4 x\, \bar{\psi}_{p'\lambda'}\gamma_{\mu}\psi_{p\lambda} \left({\mathcal A}^{\mu}\right)^*
\eea
where the final electron is also in the Volkov state,
\bea
& \bar{\psi}_{p'\lambda'}(\br,t) = N_{e'}\, \bar{u}'\, \left(1 + \fr{e}{2(p'k)}(\gamma A) (\gamma k)\right)\,\exp\left\{ip'x + \fr{ie}{(p'k)} \int\limits^{kx}d\varphi \left((p'A) - \fr{e}{2}A^2\right)\right\}.
\eea
where $\bar{u}'u' = 2m_e$.

\begin{figure}[t]
	\hspace*{-0.2cm}
	\center
	\includegraphics[width=0.6\linewidth]{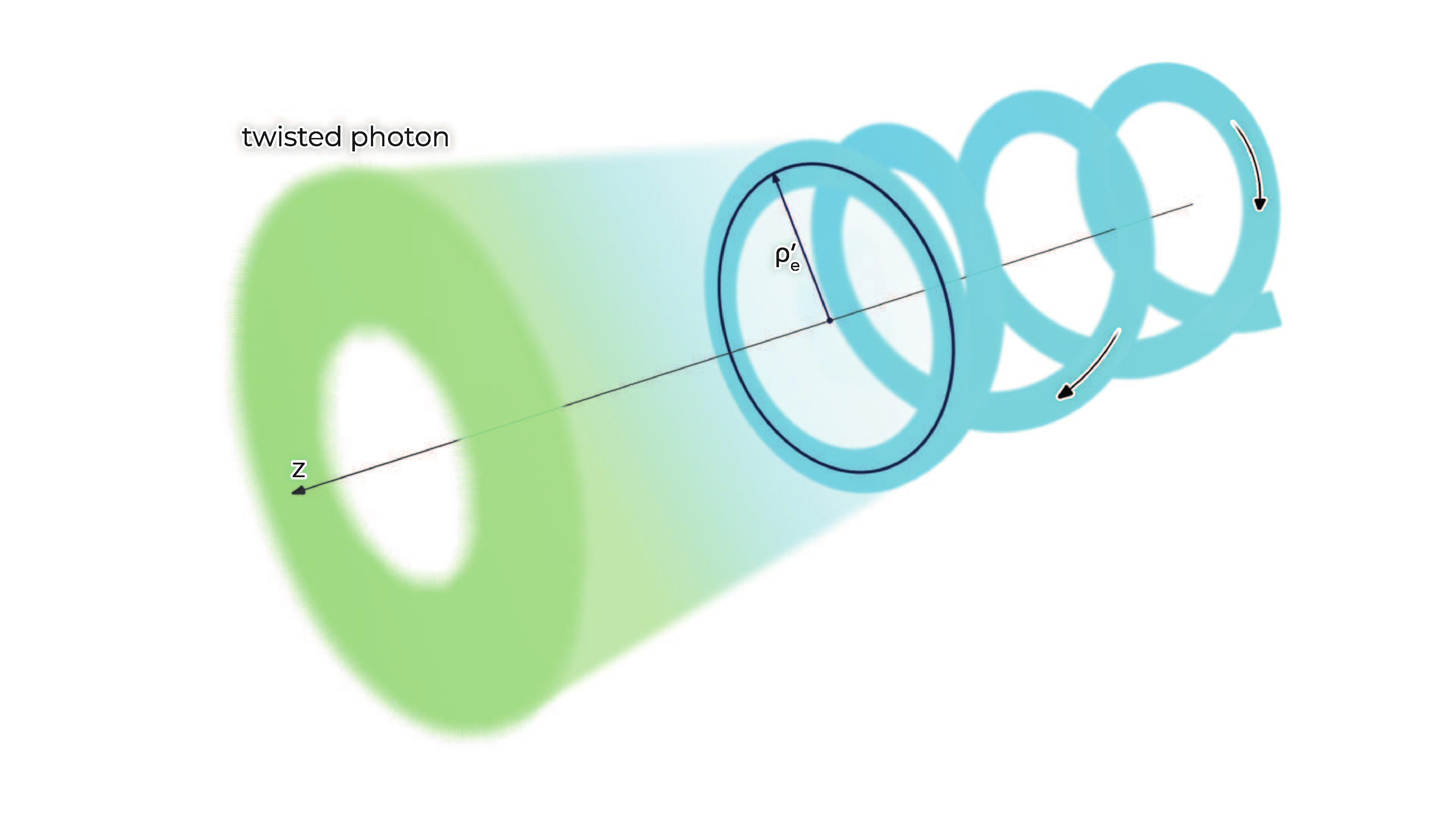}
	\caption{A schematic picture of emission of a photon during the motion of an electron wave packet along a helical path with a mean radius $\rho'_e = ea/(p'k)$, either in a helical undulator or in a circularly polarized laser wave. The evolved state of the photon becomes twisted only if the azimuthal angle of the emitting electron momentum is measured with a large error or is not measured at all, which is typical for the quasi-classical regime with the small scattering angle, $\theta' \ll 1$. The radius of the first Bessel ring is of the order of $(p'_{\perp})^{-1}$, which is usually larger than $\rho'_e$ according to Eq.(\ref{Radineq}).
\label{Fig5}}
\end{figure}

As this problem can be solved exactly \cite{BLP, R, Piazza, Fedotov}, its results can be applied for an approximate description of similar problems where an electron also moves along a helical path, the simplest example being emission in a helical undulator \cite{Ban, UndCompt, Halav}. As shown, for instance, in Ref.\cite{UndCompt}, the spectral distributions of the radiated energy of both the processes are quantitatively very similar for ultrarelativistic electrons with $\varepsilon/m \gg 1$ and the small recoil $\omega'/\varepsilon \ll 1$. We emphasize that although the electron is usually post-selected to the Volkov (plane-wave) state in calculations \cite{BLP, R, Piazza, Fedotov}, this implies \textit{a projective measurement} of the electron quasi-momentum $q'$ with the vanishing uncertainties of all its components (the scheme \textit{(i)} from Sec.\ref{Sch}).

Importantly, this may \textit{not} necessarily happen in practice if we do not have complete information about the final electron state. Indeed, when the recoil is small and the electron scattering angle is vanishing, $\theta' \ll 1$, the electron stays in the laser field or inside an undulator, its energy is thereby implicitly measured, but the momentum azimuthal angle may not be exactly known, simply because it is practically challenging to precisely measure it at the angles of $\theta' \sim 10^{-6} - 10^{-5}$ rad. Finally, inside a long undulator the number of emitted photons is large but the final electron azimuthal angle is not measured after each emission event at all. In this regime, the electron quasi-momentum is measured within the scheme \textit{(iii)} from Sec.\ref{Sch}, and the photon itself is automatically projected to the twisted state (see Fig.\ref{Fig5}). We will return to this generalized-measurement scheme in the next sub-section.

Following the standard procedure \cite{BLP, R, Piazza, Fedotov}, we expand the matrix element into series over the harmonic number $s \geqslant 1$, regroup the terms in the pre-exponential factor with the same indices of the Bessel functions $J_s$ and $J_{s\pm 1}$, and represent the result as follows:
\bea
& S_{fi}^{(\text{pw})} = \sum\limits_{s=1}^{\infty} S_{fi}^{(s)} = -ie N \sum\limits_{s=1}^{\infty} (2\pi)^4 \delta^{(4)} (q + sk - q'- k') \sum\limits_{\sigma = 0, \pm 1} (e'_{\mu})^*\,\bar{u}'\,\Gamma_{\sigma}^{\mu}\,u\, J_{s+\sigma}(\Sigma)\, e^{i (s + \sigma)\xi}.
\label{SVolkov}
\eea
Here, $N = N_e N_{e'} N_{\gamma}$ is a normalization constant, the electron quasi-momenta are
\bea
& q^{\mu} = p^{\mu} + e^2\,k^{\mu}\fr{a^2}{2(pk)},\, (q')^{\mu} = (p')^{\mu} + e^2\,k^{\mu}\fr{a^2}{2(p'k)},
\eea
and we have also denoted
\bea
& \Sigma = e \sqrt{\left(\fr{(pa_1)}{(pk)} - \fr{(p'a_1)}{(p'k)}\right)^2 + \left(\fr{(pa_2)}{(pk)} - \fr{(p'a_2)}{(p'k)}\right)^2},\cr
& \xi = \arctan\fr{(pa_2)/(pk) - (p'a_2)/(p'k)}{(pa_1)/(pk) - (p'a_1)/(p'k)}.
\eea
The ``dressed'' vertex $\Gamma_{\sigma}^{\mu}$ is
\bea
& \Gamma_{\sigma}^{\mu} = \{\Gamma_{0}^{\mu}, \Gamma_{+1}^{\mu}, \Gamma_{-1}^{\mu}\} = \cr
& = \Big\{\gamma^{\mu} - k^{\mu} \fr{e^2 A^2}{2(pk)(p'k)} (\gamma k), \fr{1}{2}(\gamma a_-) \left (\fr{e}{2} (\gamma k) \gamma^{\mu} (\fr{1}{(p'k)} - \fr{1}{(pk)}) + k^{\mu} \fr{e}{(pk)}\right) - a_-^{\mu}\fr{e}{2(pk)} (\gamma k),\cr & \fr{1}{2}(\gamma a_+) \left (\fr{e}{2} (\gamma k) \gamma^{\mu} (\fr{1}{(p'k)} - \fr{1}{(pk)}) + k^{\mu} \fr{e}{(pk)}\right) - a_+^{\mu}\fr{e}{2(pk)} (\gamma k)\Big\},\cr
& a_{\pm}^{\mu} = a_1^{\mu} \pm i a_2^{\mu}, (ka_{\pm}) = 0.
\label{VVertex}
\eea
These expressions are still the standard ones \cite{BLP, R, Piazza, Fedotov}, just written differently.

In what follows, we only study the head-on collision, for which
\bea
& p = \{\varepsilon, 0, 0, |\bp|\},\ k = \{\omega, 0,0,-\omega\},\ k'= \{\omega', k'_{\perp}  \cos\phi_{k'}, k'_{\perp} \sin\phi_{k'}, k'_z\},\,k'_{\perp} = \omega'\sin\theta_{k'},\cr 
& p'= \{\varepsilon', p'_{\perp}\cos\phi', p'_{\perp}\sin\phi', p'_z\},\ p'_{\perp} = |\bp'| \sin\theta',\cr
& a_1^{\mu} = a \{0,1,0,0\}, a_2^{\mu} = a \{0,0,1,0\},\cr
& a_{\pm}^{\mu} = a \{0,{\bm e}_{\pm}\},\ {\bm e}_{\pm} = \hat{{\bm x}} \pm i \hat{{\bm y}},\ \hat{{\bm x}} = \{1,0,0\},\ \hat{{\bm y}} = \{0,1,0\},\cr
& \delta^{(4)} (q + sk - q'- k') = \delta (q^0 + s\omega - (q')^0- \omega')\delta (q_z + sk_z - q_z'- k_z') \delta^{(2)}(\bp'_{\perp} + \bk'_{\perp}),
\label{HO}
\eea
and the transverse delta-function can again be represented in cylindrical coordinates, see Eq.(\ref{deltatr}).
In this case, $(pa_1) = (pa_2) = 0$, the vertex $\Gamma_{\sigma}^{\mu}$ does not depend on the azimuthal angles of the final particles, and so
\bea
\tan\xi = \fr{p_y'}{p_x'} = \tan\phi' \Rightarrow \xi = \phi' + \pi q,\ q = 0, 1, 2,...
\eea
For $q=0$, we have
\be
\xi = \phi' = \phi_{k'} \pm \pi
\ee
due to the transverse delta-function. The argument of the Bessel functions becomes
\bea
\Sigma = \fr{ea}{(p'k)}p'_{\perp} = \fr{ea}{(p'k)}k'_{\perp} \equiv \rho'_e k'_{\perp},
\eea
where 
\bea
& \rho'_e = \fr{ea}{(p'k)} = \eta\, \fr{m_e}{(p'k)}
\label{rhoe}
\eea
is a classical radius of the final electron path (see the problem No.2 to Sec.47 in Ref.\cite{LL2} and Fig.\ref{Fig5}) and
\bea
& \eta = \fr{e\sqrt{-A^2}}{m_e} = \fr{ea}{m_e}
\label{Eta}
\eea
is a classical field strength parameter \cite{BLP, R, Piazza, Fedotov}. Thus, the radius of the classical helical trajectory inside the laser wave \textit{naturally arises} in the quantum theory.
For an optical photon and a relativistic electron, we have roughly
\bea
& \rho'_e \sim \eta\,\lambda_c\, \fr{m_e}{\varepsilon'} \fr{m_e}{\omega} \sim \fr{\eta}{\gamma'}\times 0.1 [\mu\text{m}],\cr 
& \gamma' = \varepsilon'/m_e,\ \lambda_c = 1/m_e,
\eea
which reaches the atomic scale, 
\be
\rho'_e \sim 0.1\, \text{nm} - 1\, \text{nm},
\ee
for $\gamma' \sim 10-100, \eta \sim 0.1-1$.

Thus, the evolved state of the emitted photon (\ref{EvWV}) looks as follows:
\bea
& {\bm A}^{(\text{ev})}(\bk', \omega') = \sum\limits_{s=1}^{\infty} {\bm A}^{(\text{ev},s)}(\bk', \omega'),\cr 
& \bk'\cdot{\bm A}^{(\text{ev},s)} = 0,\ {\bm n}' = \bk'/\omega',\cr  
& {\bm A}^{(\text{ev},s)}(\bk', \omega') = -ie N (2\pi)^4 \delta^{(4)} (q + sk - q'- k') \cr
& \times \hspace{-5pt} \sum\limits_{\lambda_{\gamma}=\pm 1}\sum\limits_{\sigma=0,\pm 1} {\bm e}' e'^*_{\mu}\,\bar{u}'\,\Gamma_{\sigma}^{\mu}\,u\, J_{s+\sigma}(\rho'_e k'_{\perp}) \, e^{i (s + \sigma)\phi'},
\label{Aevs}
\eea
where the sum over $\sigma$ has appeared because of the spin terms (\ref{spin}). Summation over the photon helicities $\lambda_{\gamma}$ is done as in Eq.(\ref{sumhel}). Then, choosing the final electron bispinor as an eigenstate of $\hat{j}'_z$ operator with the eigenvalue $\lambda' = \pm 1/2$, as in Sec.\ref{Phases}, the evolved state becomes
\bea
& {\bm A}^{(\text{ev},s)}(\bk', \omega') = -ie N (2\pi)^4 \delta^{(4)} (q + sk - q'- k')\cr 
& \times {\bm n}' \times \left[{\bm n}' \times \sum\limits_{\sigma' = \pm 1/2}\sum\limits_{\sigma=0,\pm 1} d^{(1/2)}_{\sigma'\lambda'}(\theta')\, \bar{u}_{\varepsilon'\lambda'}^{(\sigma')}\,{\bm \Gamma}_{\sigma}\,u\, J_{s+\sigma}(\rho'_e k'_{\perp})\, e^{i (s + \sigma + \sigma' - \lambda')\phi'}\right].
\label{Astr}
\eea
where we have expanded the final electron bispinor according to Eq.(\ref{uexplambda}). Clearly, this wave function is just a plane wave with the vanishing $z$-projection of the OAM because the photon has the definite 4-momentum $k'$ and the azimuthal angle $\phi_{k'}$.

\subsection{Post-selection with a generalized measurement}\label{ComptGen}
\subsubsection{Volkov electron}

In the standard projective-measurement scheme, we detect photons emitted in a laser wave or in a helical undulator with a detector placed at an angle $\phi_{k'}$, which automatically projects the electron to the plane-wave (Volkov) state $|{\bm q}',\lambda'\ra$ with $\phi' = \phi_{k'} \pm \pi$. However, if we do not measure the electron angle $\phi'$ alone, as we argued above, then the photon itself evolves to the twisted state, irrespectively of what detector is subsequently used. The electron detected state in the generalized-measurement scheme with the maximized error is 
\be
|e'_{\text{det}}\ra^{(\text{g})} = \int\limits_0^{2\pi} \fr{d\phi'}{2\pi}\,|{\bm q}',\lambda'\ra,
\ee
and the information about the electron quasi-momentum ${\bm q}'$ is incomplete. Importantly, in the head-on geometry (\ref{HO}) we have $\bp'_{\perp} = {\bm q}'_{\perp}$, and so the azimuthal angles of both the vectors \textit{coincide}.

We can obtain the evolved photon wave function by integrating Eq.(\ref{Astr})
\bea
& {\bm A}^{(\text{ev},s)}_{(\text{g})}(\bk', \omega') = \int\limits_0^{2\pi}\fr{d\phi'}{2\pi}\, {\bm A}^{(\text{ev},s)}(\bk', \omega') = \cr & = -ie N (2\pi)^3 \delta (q^0 + s\omega - (q')^0- \omega') \delta (q_z + sk_z - q_z'- k_z')\,\fr{1}{k_{\perp}'}\delta (p_{\perp}' - k_{\perp}')\cr 
& \times {\bm n}' \times \left[{\bm n}' \times \sum\limits_{\sigma' = \pm 1/2}\sum\limits_{\sigma=0,\pm 1} d^{(1/2)}_{\sigma'\lambda'}(\theta')\, \bar{u}_{\varepsilon'\lambda'}^{(\sigma')}\,{\bm \Gamma}_{\sigma}\,u\, J_{s+\sigma}(\rho'_e k'_{\perp})\, e^{i (s + \sigma + \sigma' - \lambda')(\phi_{k'} \pm \pi)}\right].
\label{Acylpol}
\eea
Even without the calculation of $\bar{u}_{\varepsilon'\lambda'}^{(\sigma')}\, {\bm \Gamma}_{\sigma}\,u$, it is now clear that the photon represents a cylindrical wave with the SOI. 

Next, we use the representation 
\be
\bar{u}_{\varepsilon'\lambda'}^{(\sigma')}\,{\bm \Gamma}_{\sigma}\,u = {\bm G}^{(\uparrow \uparrow)}\, \delta_{\lambda,\sigma'} + {\bm G}^{(\uparrow \downarrow)}\, \delta_{\lambda,-\sigma'},
\ee
where the vectors ${\bm G}^{(\uparrow \uparrow)} \equiv {\bm G}^{(\uparrow \uparrow)}_{\sigma\lambda'\lambda}, {\bm G}^{(\uparrow \downarrow)} \equiv {\bm G}^{(\uparrow \downarrow)}_{\sigma\lambda'\lambda}$ are found as
\bea
&& {\bm G}^{(\uparrow \uparrow)}_{0\lambda'\lambda} = {\bm \chi}_{0} \left( f^{(2)}_{\lambda'\lambda} - \fr{\eta^2 m_e^2 \omega^2}{2(pk)(p'k)} (f^{(1)}_{\lambda'\lambda} + f^{(2)}_{\lambda'\lambda})\right),\cr
&& {\bm G}^{(\uparrow \downarrow)}_{0\lambda'\lambda} = - f^{(2)}_{\lambda'\lambda} \sqrt{2}\, {\bm \chi}_{2\lambda},\cr
&& {\bm G}^{(\uparrow \uparrow)}_{\pm 1\lambda'\lambda} = \mp\sqrt{2}\, {\bm \chi}_{\mp 1} \fr{\eta m_e \omega}{2} \left(f^{(1)}_{\lambda'\lambda} + f^{(2)}_{\lambda'\lambda}\right) \left(\delta_{\lambda,\mp 1/2} \left(\fr{1}{(p'k)} - \fr{1}{(pk)}\right) + \fr{1}{(pk)}\right),\cr
&& {\bm G}^{(\uparrow \downarrow)}_{\pm 1\lambda'\lambda} = \mp {\bm \chi}_{0} \fr{\eta m_e \omega}{2}\delta_{\lambda,\pm 1/2} \left(\left(\fr{1}{(p'k)} - \fr{1}{(pk)}\right) (f^{(1)}_{\lambda'\lambda} - f^{(2)}_{\lambda'\lambda}) - \fr{2f^{(2)}_{\lambda'\lambda}}{(pk)}\right),\cr
&& f^{(1)}_{\lambda'\lambda} = \sqrt{\varepsilon + m_e} \sqrt{\varepsilon' + m_e} + 2\lambda 2\lambda' \sqrt{\varepsilon - m_e} \sqrt{\varepsilon' - m_e},\cr
&& f^{(2)}_{\lambda'\lambda} = \sqrt{\varepsilon - m_e} \sqrt{\varepsilon' + m_e} + 2\lambda 2\lambda' \sqrt{\varepsilon + m_e} \sqrt{\varepsilon' - m_e},
\label{GG}
\eea
and they do not explicitly depend on the photon 4-momentum $k'$. The vectors ${\bm \chi}_{0}, {\bm \chi}_{\pm 1}$ are given in Eq.(\ref{chiv}). The final expression for the wave function is
\bea
& {\bm A}^{(\text{ev},s)}_{(\text{g})}(\bk', \omega') = -ie (-1)^{s+\lambda-\lambda'} N (2\pi)^3 \delta (q^0 + s\omega - (q')^0- \omega') \delta (q_z + sk_z - q_z'- k_z')\,\fr{1}{k_{\perp}'}\delta (p_{\perp}' - k_{\perp}')\cr 
& \times {\bm n}' \times \left[{\bm n}' \times \sum\limits_{\sigma=0,\pm 1} J_{s+\sigma}(\rho'_e k'_{\perp})\, e^{i (s + \sigma - \lambda')\phi_{k'}} \left(d^{(1/2)}_{\lambda\lambda'}(\theta')\,{\bm G}_{\sigma\lambda'\lambda}^{(\uparrow \uparrow)}\, e^{i\lambda\phi_{k'}} +
d^{(1/2)}_{-\lambda\lambda'}(\theta')\,{\bm G}_{\sigma\lambda'\lambda}^{(\uparrow \downarrow)}\, e^{-i\lambda\phi_{k'}}\right)\right],
\label{Acylpol2}
\eea
which is \textit{a Bessel beam} with the SOI, revealed by the sum over $\sigma$, and the following TAM projection:
\bea
\hat{j}_z^{(\gamma)} {\bm A}^{(\text{ev},s)}_{(\text{g})} = (s + \lambda - \lambda') {\bm A}^{(\text{ev},s)}_{(\text{g})},
\eea
whereas the TAM uncertainty is vanishing. Here, in contrast to Eq.(\ref{TAMcl1}) for Cherenkov radiation, the harmonic number $s=1,2,3,...$ appears in the r.h.s. 
For the unpolarized electrons with $\la \hat{j}_z\ra = \la \hat{j}'_z\ra = 0$, the photon TAM is simply $s$. Analogously to Cherenkov radiation, the transverse momentum of this Bessel photon is defined by the electron scattering angle $\theta'$ as
\be
k'_{\perp} = |\bp'| \sin\theta'.
\ee

As argued in Sec.\ref{EvSt}, the coordinate representation for 1-particle evolved states has a clear interpretation only when the other particle wave-packet is localized spatially and temporarily, and so the electron detector function $f_p$ is smooth. Nevertheless -- even without such a localization -- the monochromatic Bessel beam in the coordinate representation
\bea
& \displaystyle {\bm A}^{({\text{ev}},s)}_{(\text{g})}(\br,t) = \int\frac{d^3k'}{(2\pi)^3}\,\frac{1}{\sqrt{2\omega'}}\,{\bm A}^{(\text{ev}),s}_{(\text{g})}(\bk',\omega')\, e^{-i\omega't + i\bk'\cdot\br}
\eea
can be used to visualize spatial distribution of the emitted energy. To this end, it is convenient to divide the vector potential (\ref{Acylpol2}) into the following parts:
\bea
&{\bm A}^{({\text{ev}},s)}_{(\text{g})} \equiv {\bm A}^{({\text{ev}},s)}_{\bm G} + {\bm A}^{({\text{ev}},s)}_{{\bm n}'},\cr
& {\bm A}^{({\text{ev}},s)}_{\bm G} \propto {\bm G},\ {\bm A}^{({\text{ev}},s)}_{{\bm n}'} \propto {\bm n}'({\bm n}' \cdot {\bm G}), 
\label{Furier}
\eea
where ${\bm G}$ is either ${\bm G}_{\sigma\lambda'\lambda}^{(\uparrow \uparrow)}$ or ${\bm G}_{\sigma\lambda'\lambda}^{(\uparrow \downarrow)}$. For the former part, we obtain \footnote{Recall that we have not used any localizing function $f_p$ for the electron from Eq.(\ref{fp}), which is why ${\bm A}^{({\text{ev}},s)}_{(\text{g})}(\br,t)$ stays proportional to the delta-function. The regularization for this delta-function squared in Eq.(\ref{Espat}) is analogous to the use of some smooth function $f_p$.}
\bea
& \displaystyle {\bm A}^{({\text{ev}},s)}_{\bm G}(\br,t) = ie (-1)^{s+\lambda-\lambda'} N\,\frac{2\pi}{\sqrt{2\omega'}}\,\delta(q^0 + s\omega - (q')^0 - \omega')\, e^{-it(q^0 + s\omega - (q')^0) + iz(q_z + sk_z - q_z')}\cr & \times \sum\limits_{\sigma=0,\pm 1} i^{s + \sigma - \lambda'}\, J_{s+\sigma}(\rho'_e p'_{\perp})\, e^{i (s + \sigma - \lambda')\phi_r} \Big(i^{\lambda}\, d^{(1/2)}_{\lambda\lambda'}(\theta')\,{\bm G}_{\sigma\lambda'\lambda}^{(\uparrow \uparrow)}\, e^{i\lambda\phi_r}\, J_{s + \sigma -\lambda' + \lambda}(\rho p'_{\perp}) + \cr
& + i^{-\lambda} d^{(1/2)}_{-\lambda\lambda'}(\theta')\,{\bm G}_{\sigma\lambda'\lambda}^{(\uparrow \downarrow)}\, e^{-i\lambda\phi_r}J_{s + \sigma -\lambda' - \lambda}(\rho p'_{\perp})\Big),
\label{Acylpol3}
\eea
where $\omega' = q^0 + s\omega - (q')^0 = \sqrt{(p'_{\perp})^2 + (q_z + sk_z - q_z')^2}$ and $\phi_r$ is the azimuthal angle of the vector 
\be
\br = \{\rho \cos\phi_r, \rho \sin\phi_r, z\}, 
\ee
and the following identity has been used
\bea
\int\limits_0^{2\pi}\frac{d\phi}{2\pi}\,e^{ix\cos\phi + i\ell\phi} = i^{\ell} J_{\ell}(x).
\eea
Clearly, the part ${\bm A}^{({\text{ev}},s)}_{{\bm n}'}(\br,t)$ is much more cumbersome. 

\begin{figure*}[t]
	\hspace*{-0.2cm}
	\center
	\includegraphics[width=1.0\linewidth]{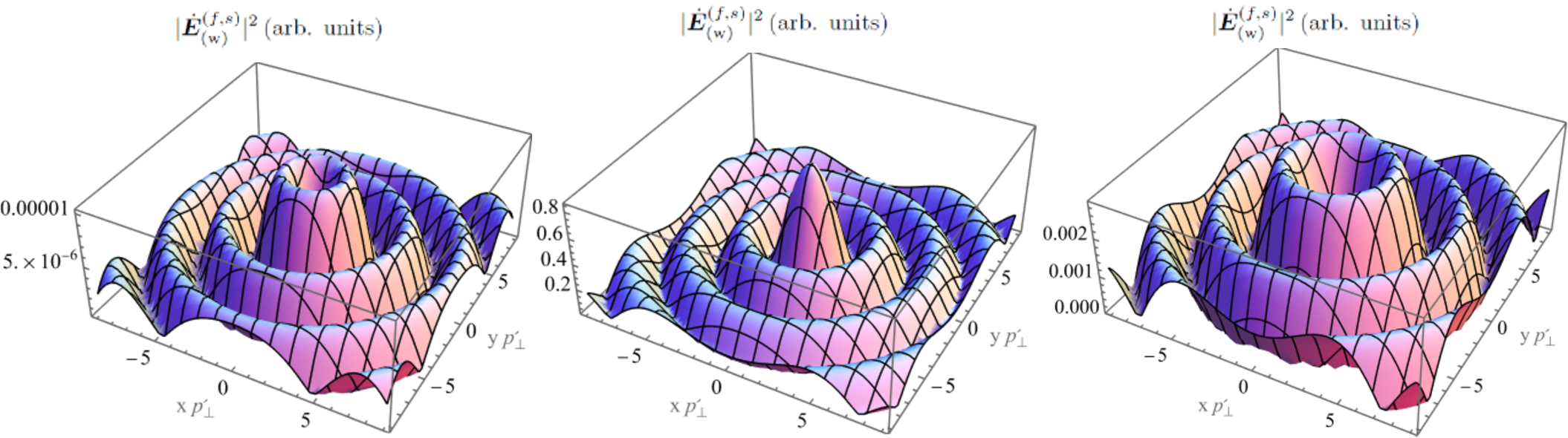}
	\caption{The regularized spatial energy distributions (\ref{Espat}) of the Bessel photon emitted in the non-linear Compton scattering within the generalized-measurement scheme \textit{(iii)} with the electron azimuthal angle being measured with a very large error or not measured at all. Left panel: $\omega = 5\,\text{eV},\, \varepsilon/m_e = 10, \varepsilon' = 0.9999\, \varepsilon, \theta' \approx 10^{-5}\, \text{rad}, \eta = 0.5, s = 1, \lambda = -1/2, \lambda' = 1/2$ (a spin flip), $j_z^{(\gamma)} = s +\lambda - \lambda' = 0$. Central panel: the same, but $\lambda = 1/2, \lambda' = 1/2$ (no spin flip), $j_z^{(\gamma)} = s +\lambda - \lambda' = 1$. Right panel: $\omega = 0.5\,\text{eV},\, \varepsilon/m_e = 100, \varepsilon' = 0.9999\, \varepsilon, \theta' \approx 10^{-6}\, \text{rad}, \eta = 0.5, s = 3, \lambda = 1/2, \lambda' = 1/2$ (no spin flip), $j_z^{(\gamma)} = s +\lambda - \lambda' = 3$. Everywhere $(p'_{\perp})^{-1} \approx 4\, \text{nm} \approx 4 \rho'_e$, so the Bessel rings are wider than the radius of the electron classical path $\rho'_e$ (\ref{rhoe}) according to Eq.(\ref{Radineq}).
\label{Fig6}}
\end{figure*}

The electric field strength is ${\bm E}_{(\text{g})}^{({\text{ev}},s)}(\br, t) = -\partial {\bm A}^{({\text{ev}},s)}_{(\text{g})}(\br,t)/\partial t$, and the regularized spatial distributions of the energy of this Bessel beam 
\bea
|\dot{{\bm E}}_{(\text{g})}^{({\text{ev}},s)}|^2 = \frac{1}{T}\,\int d\omega' |{\bm E}_{(\text{g})}^{({\text{ev}},s)}(\br, t)|^2 
\label{Espat}
\eea
represent a typical doughnut, as shown in Fig.\ref{Fig6}. Here $T$ is a very large ``observation time'' obtained when squaring the delta-function in Eq.(\ref{Acylpol3}) (see, e.g., \cite{BLP}). Importantly, whereas the radius of the emitting electron classical path (\ref{rhoe}) does not change much during the emission of soft photons, $\rho'_e \approx \rho_e$, the radius of the first Bessel ring is of the order of $1/p'_{\perp} \approx 1/\varepsilon'\theta'$, which \textit{does not necessarily coincide} with $\rho_e$. Indeed, the electron scattering angle is usually small and limited from above \cite{VG2004},
\bea
& \displaystyle \theta' \leq \frac{2s}{\sqrt{1 + \eta^2}}\, \frac{\omega}{m_e},
\eea
so the photon transverse momentum is
\bea
& \displaystyle k'_{\perp} \leq \varepsilon'\, \frac{2s}{\sqrt{1 + \eta^2}}\, \frac{\omega}{m_e},
\label{thIneq}
\eea
and it can reach the keV scale for GeV electrons and optical laser photons. Detailed study of the quasi-classical regime of emission by ultrarelativistic electrons is presented in the Appendix, Sec.\ref{QC}, and comparison with the predictions of classical electrodynamics are given in the Appendix, Sec.\ref{AppB}. That is why in the relativistic case we have
\bea
& \displaystyle \frac{(p'_{\perp})^{-1}}{\rho'_e} \approx \frac{2}{\eta} \frac{\omega}{m_e} \frac{1}{\theta'} \geq \frac{\sqrt{1 + \eta^2}}{s\, \eta}.
\label{Radineq}
\eea
Very roughly, one can think of $(k'_{\perp})^{-1} = (p'_{\perp})^{-1}$ as of \textit{the transverse coherence length} of the twisted photon, although this quantity cannot be quantitatively defined for an unnormalized Bessel beam. Thus, in the linear regime with $\eta \ll 1$ the radius of the first Bessel ring is much larger than the radius of the electron classical path, while in the non-linear regime with $\eta \gtrsim 1$ this ratio is of the order of unity, as shown in Fig.\ref{Fig5}. The opposite case with $\eta \gg 1$ corresponds to a constant crossed field instead of the plane wave \cite{BLP, R}, and the above ratio gets larger than $1/s$.










\subsubsection{Bessel-Volkov electron}

Let us now turn to the case when the incoming electron is twisted with the TAM $m = \pm 1/2, 3/2,...$ and it is in the Bessel-Volkov state \cite{PRA12}. 
Clearly, this time we obtain 
\bea
\hat{j}_z^{(\gamma)} {\bm A}^{(f,s)}_{(\text{g})} = (s + m - \lambda') {\bm A}^{(f,s)}_{(\text{g})},
\eea
whereas the most general formula for the mixed states is
\be
\la \hat{j}_z\ra^{(\gamma)} = s + \la \hat{j}_z\ra - \la \hat{j}'_z\ra.
\ee

Thus, highly twisted photons with $\la \hat{j}_z^{(\gamma)}\ra \gg 1$ can be obtained via the non-linear Compton effect 
\begin{itemize}
\item
Either by using the plane-wave (Volkov) incoming electrons and a powerful laser with $\eta \gtrsim 1$, so that the higher harmonics with $s > 1$ are generated,
\item
Or, alternatively, by taking a {\it moderately powerful} laser, for which only the linear scattering with $s=1$ takes place, 
but with an incoming vortex (Bessel-Volkov) electron with $|m| \gg 1$. 
\end{itemize}
As the highly twisted electrons with $m \sim 100-1000$ have already been obtained \cite{Bliokh}, the latter scheme seems to be technically easier. 
Besides, if the vortex electron is ultrarelativistic, the energy of the scattered highly twisted photons can easily reach MeV range \cite{Budker}. The means for accelerating the vortex electrons to ultrarelativistic energies have been recently discussed in Ref.\cite{NJP}.

We emphasize that in the quantum regime the twisted photons are \textit{not naturally generated} in a circularly-polarized laser wave or in a helical undulator, as it seems from the classical theory \cite{Sasaki1, Sasaki2, Afan, Taira, Katoh, Katoh2}. In quantum approach, the photon turns out to be twisted only if the measurement error of the azimuthal angle of the final electron momentum is large. When the recoil angle $\theta'$ is small, the angle $\phi'$ simply loses its sense, which is why this generalized-measurement scheme reproduces results of the classical theory.

\section{Example 3: Heavy and composite particles} 
\subsection{Leptons}

An advantage of the generalized-measurement technique is that it can in principle be applied to particles of any mass, spin, and energy, including composite particles. 
We start with the QED scattering with a lepton heavier than electron, 
\be
e^{\pm}\mu^{\pm} \to e^{\pm}\mu^{\pm}\quad \text{or}\quad e^{\pm}\tau^{\pm} \to e^{\pm}\tau^{\pm}.
\ee 
For definiteness we take the process
 \be
e^{-}(p_1)+\mu^{-}(p_2) \to e^{-}(p_3) +\mu^{-}(p_4) 
 \ee
as an example, where the electron and the muon collide head-on and have the momenta 
\bea
& p_1 = \{\varepsilon_1, 0, 0,|\bm p_1|\},\, \varepsilon_1 = \sqrt{m_e^2 + |\bm p_1|^2},\cr 
& \text{and}\quad p_2 = \{\varepsilon_2 , 0, 0, -|\bm p_2|\},\, \varepsilon_2 = \sqrt{m_{\mu}^2 + |\bm p_2|^2},
\eea
as well as the helicities $\lambda_1, \lambda_2$, respectively. 
When the final electron is detected in the generalized-measurement scheme, the evolved wave function of the muon, according to Eq.(\ref{EvWV}), becomes
\be
\psi^{(f)}_\mu = \sum\limits_{\lambda_4 = \pm 1/2} 
\int\limits_0^{2\pi} \fr{d\phi_3}{2\pi}\,
u_4\, S_{fi}^{\rm(pw)},
\ee
where $u_4 \equiv u_{p_4\lambda_4}$ and the matrix element reads
\bea
& S_{fi}^{\rm(pw)} = i(2\pi)^4 N\, \delta^{(4)}(p_1 + p_2 - p_3 - p_4)\cr 
& \times \fr{4\pi e^2}{q^2}\, \left(\bar{u}_3 \gamma^{\alpha}u_1\right)\, \left(\bar{u}_4 \gamma_{\alpha}u_2\right),\, q = p_1 - p_3,
\eea
where the photon propagator is taken in the Feynman gauge. In contrast to the previous section, the 4-vector $q$ is a momentum transfer. 

\begin{figure*}[t]
	\center
	\includegraphics[width=0.8\linewidth]{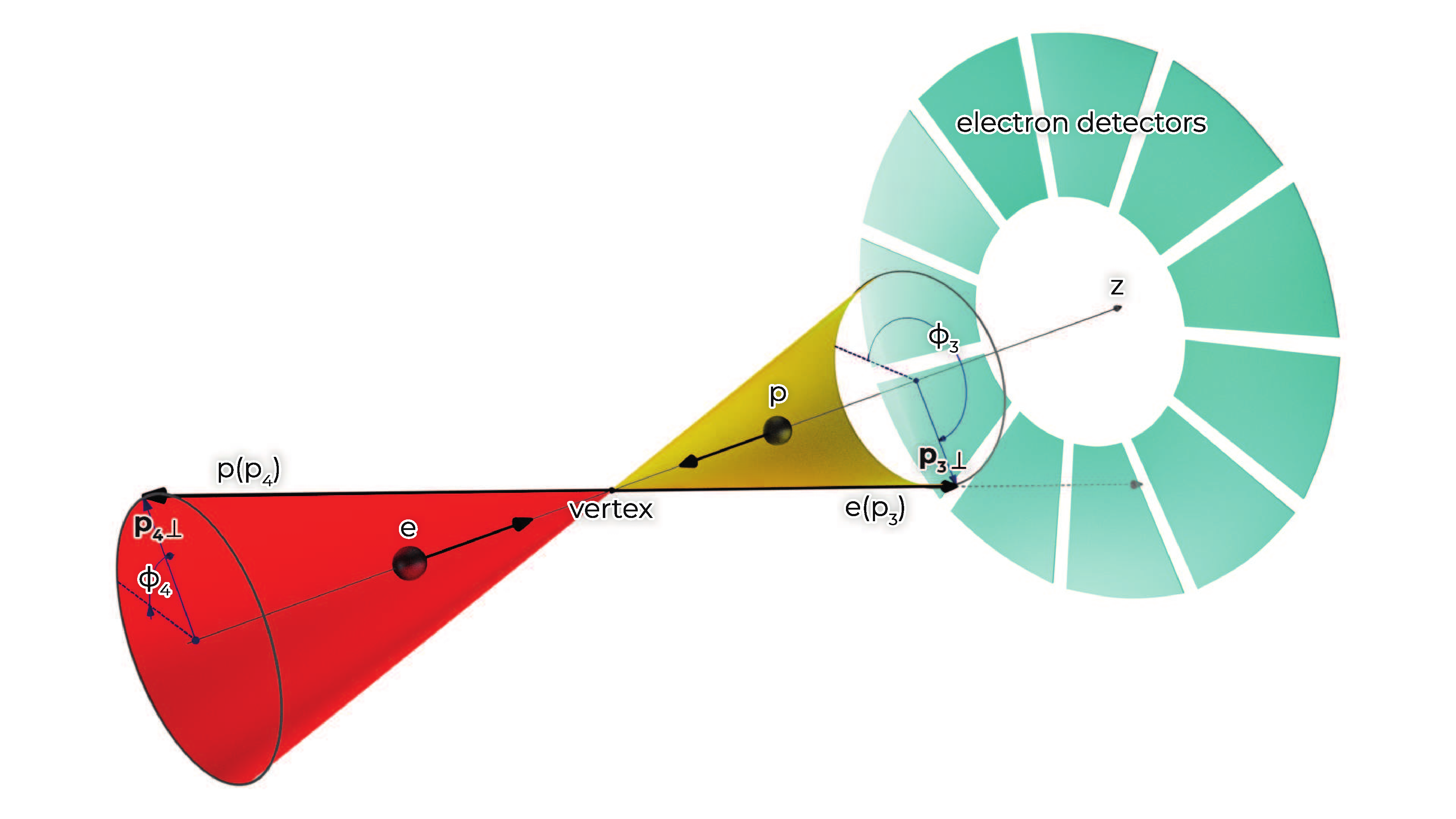}
	\caption{An azimuthal ``which-way'' experiment with elastic $ep \to ep$ scattering. The evolved state of the proton becomes twisted if the electron azimuthal angle $\phi_3$ is measured with a large error or not measured at all. The same scheme can also be applied to elastic and inelastic processes with other hadrons, ions, and nuclei.
\label{Fig7}}
\end{figure*}

The muon current can be presented as 
\bea
\bar{u}_4 \gamma_{\alpha} u_2 = J^{(\mu)}_{\alpha} e^{-i(\lambda_2+\lambda_4)\phi_4},
\eea
where $J^{(\mu)}_{\alpha}$ is given in Eq.(\ref{jopp}) of the Appendix \ref{Phases} with $m_e \to m_{\mu}$. 
After the integration, the electron current becomes
\bea
& \int\limits_0^{2\pi} d\phi_3\,\delta (\phi_3 - (\phi_4 \pm \pi))\, \bar{u}_3 \gamma_{\alpha} u_1 = \cr
& = J^{(e)}_{\alpha} (\phi_3 = \phi_4 \pm \pi) e^{i(\lambda_1-\lambda_3)(\phi_4 \pm \pi)}.
\label{Je}
\eea
Taking into account that 
\bea
{\bm \chi}_{-2\lambda_2}\cdot{\bm \chi}_{2\lambda_1} = -\delta_{\lambda_2 \lambda_1},
\eea
it is easy to check that the scalar product
\bea
\mathcal J \equiv J_{\alpha}^{(e)}(\phi_3 = \phi_4 \pm \pi)\, (J^{(\mu)})^{\alpha}
\eea 
does not depend on $\phi_4$. Therefore,  
 \bea
 \psi^{(f)}_\mu \propto \sum\limits_{\lambda_4 = \pm 1/2} \tilde{u}_4\,\, \mathcal J\, e^{i(\lambda_1-\lambda_2-\lambda_3)\phi_4},
 \eea
where $\tilde{u}_4 = u_4\,e^{- i \lambda_4 \phi_4}$ has a vanishing TAM, see Eq.(\ref{jz0}) in the Appendix \ref{Phases}. As a result, 
 \be
{\hat j}_{4,z}\psi^{(f)}_\mu= (\lambda_1-\lambda_2-\lambda_3)\psi^{(f)}_\mu,
\ee
i.e. the muon becomes twisted. 

Similarly, for the twisted incoming electron with $m = \pm 1/2, 3/2,\ldots$, the TAM of the final muon becomes 
\be
j_{4,z} = m -\lambda_2-\lambda_3.
\ee 
Dealing with the unpolarized initial muon and the final electron, the incoming electron TAM $m$ is simply transferred to the muon,
\be
\la\hat {j}_{z}\ra^{(\mu)} = m.
\ee 

Clearly, the same TAM-transfer can be realized in other processes like $e^-e^+ \leftrightarrow 2\gamma, e^-e^+ \to \tau^-\tau^+$, etc. For instance, it has been recently shown in Ref. \cite{IvanovMu} that a hallmark of the muon's twisted state is the modification of the electron spectra as the incoming muon decays into an electron and two neutrinos. Such a modification can be experimentally observed, which would prove the vortex state of the muon.

\subsection{Hadrons}\label{Hadrons}

Finally, we study scattering off a hadron taking as an example elastic scattering of an electron by a proton with a mass $m_p$,
\be 
e(p_1) + p(p_2) \to e(p_3) + p(p_4),
\ee
in the same head-on geometry (see Fig.\ref{Fig7} \cite{Letter}) with $p_3 = \{\varepsilon_3, p_{3,\perp}\cos\phi_3, p_{3,\perp}\sin\phi_3, p_{3,z}\}$ and similarly for $p_4$. Generalization to other hadrons, ions, nuclei, or to the inelastic processes due to the non-electromagnetic forces (say, to deep-inelastic $ep$ or $p\bar{p}$ scattering) is straightforward.

The plane-wave matrix element reads
\bea
& S_{fi}^{\rm(pw)} = i(2\pi)^4 N\, \delta^{(4)}(p_1 + p_2 - p_3 - p_4)\cr 
& \times \fr{4\pi e^2}{q^2}\, \left(\bar{u}_3 \gamma^{\mu}u_1\right)\, \left(\bar{u}_4 \Gamma_{\mu}u_2\right),
\eea
where 
\bea 
\Gamma_{\mu} = F_1 \, \gamma_{\mu} + F_2 \, \sigma_{\mu\nu}q^{\nu}
\eea
is a hadronic vertex, parametrized with two form-factors \cite{BLP, Peskin} 
\be
F_1 = F_1 (q^2, P^2), F_2 = F_2 (q^2, P^2),
\ee 
where 
\be
q^2 = (p_1 - p_3)^2,\, P^2 = (p_4 + p_2)^2/4
\ee
do not depend on the angle $\phi_4$ of the final proton. So the form-factors do not depend on it either.
Here also
\bea
\sigma^{\mu\nu} = \frac{1}{2}(\gamma^{\mu}\gamma^{\nu}-\gamma^{\nu}\gamma^{\mu}) = ({\bm \alpha}, i {\bm \Sigma}),
\eea
where ${\bm \alpha}, {\bm \Sigma}$ are the $4\times 4$ Dirac matrices in the standard representation \cite{BLP}.

First, one can prove that 
\be
\bar{u}_3(\phi_3 = \phi_4 \pm \pi)\gamma^{\mu} u_1\,
\left(\bar{u}_4 \Gamma_{\mu}u_2\right)
\propto e^{i (\lambda_1 - \lambda_2 - \lambda_3 - \lambda_4)\phi_4},
\ee
and then we use Eq.(\ref{alpha}) from the Appendix \ref{Phases} for $\sigma^{\mu\nu}q_{\nu}$. 
The evolved wave function of the final proton is defined by Eq.(\ref{EvWV}),
\bea
& \displaystyle \psi^{(f)}_p = \sum\limits_{\lambda_4 = \pm 1/2} \int\limits_0^{2\pi}\fr{d\phi_3}{2\pi}\, u_4\, S_{fi}^{(pw)} 
\eea
Within the same generalized-measurement protocol (see Fig.\ref{Fig7}), the evolved wave function of the proton becomes twisted,
\be 
\hat j_{4,z} \psi^{(f)}_p= (\lambda_1-\lambda_2-\lambda_3)\, \psi^{(f)}_p.
\ee 
When the incoming electron is twisted with the TAM z-projection $m$, we have
\be
\lambda_1-\lambda_2-\lambda_3 \to m-\lambda_2-\lambda_3
\ee
If the initial hadron is unpolarized and the electron polarization is not measured, the electron TAM is transferred to the hadron, 
\be
\la\hat {j}_{z}\ra^{(p)} = m.
\ee 
Clearly, this TAM transfer can also be realized for other particles, regardless of their masses and spins, including neutrons, pions, atoms, ions, and nuclei.

\section{Discussion and conclusion}

We have theoretically demonstrated that quantum entanglement in a pair of particles together with the generalized measurement on one of them can be employed for generating a highly energetic vortex state of the other -- in principle arbitrary -- particle. 
We have given examples when the measurement error is maximized and the resultant vortex states represent the Bessel beams of photons, electrons, muons, hadrons, etc. Were this error smaller, $\sigma_{\phi} < 2\pi$, the resultant states would be the Bessel-like wave packets with \textit{a finite uncertainty} of the angular momentum but with \textit{the same central value}, defined by the TAM conservation law. 

Somewhat surprisingly, it is this inherently quantum approach with the large measurement error that most adequately describes the emission of twisted photons when the recoil is vanishing, thus providing agreement with the classical theory. More importantly, the developed model explicitly demonstrates that it is the measurement protocol that can  guarantee vorticity of a final particle, whereas helical motion of the charged particles alone (say, during the heavy-ion collisions \cite{Silenko23}) does \textit{not} automatically lead to the twisted state of the resultant particles. Indeed, even during quantum transitions between the different Landau levels of a charged particle in magnetic field the emitted photon is not always twisted \cite{HD}. 

The transverse momentum of the resultant twisted particle is defined by that of the other particle whose azimuthal angle is measured with a large error. Thus, selecting the particles scattered at larger polar angles -- say, in $ep, \gamma e, \gamma p$, $\gamma$-ion, or ion-ion collisions -- one can in principle obtain the vortex beams with the desired transverse momentum. For instance, these momenta can reach the keV scale for hard X-ray or $\gamma$-range twisted photons obtained in collisions of intense laser pulses with GeV electrons or ions (say, at the Gamma Factory \cite{Budker, Budker2}). Clearly, the number of particles scattered at large polar angles is usually not high in relativistic collisions, which puts an upper limit on the transverse momenta. Another limiting factor is the space charge, which can lead to decoherence and destroy the entanglement. 

One potential application of this generalized measurement technique can be the vorticity transfer from photons with large OAM quanta to leptons and hadrons, including ions and nuclei, for instance, via elastic scattering $\gamma l \to \gamma l$ ($l = e, \mu, \tau$), $\gamma h \to \gamma h$ ($h = p, \bar{p}, n$,...). Whereas the highest value of the photon OAM generated so far reaches $10^4$ \cite{Zeilinger10010}, the electron OAM is still one order of magnitude smaller \cite{l1000, l10002}. So one can first generate the highly twisted photons and then transfer the OAM to a massive particle. The resultant twisted matter waves can come in handy for a number of high-energy scattering and annihilation experiments, including study of the spin-OAM entanglement phenomena in particle and nuclear collisions, deep-inelastic $ep, pp, p\bar{p}$ scattering, probing spins of hadrons and nuclei, enhanced non-dipole effects in interactions with atoms and nuclei, enhanced magnetic moment contribution in interaction with surfaces and magnetic materials, channeling phenomena in crystals, and so forth.

For low-intensity beams, this method can readily be implemented for the production of highly energetic vortex particles at synchrotron radiation facilities and free-electron lasers equipped with variable-polarization  undulators like in the case of the SASE3 afterburner \cite{Kar} at the European XFEL, at the powerful laser facilities aimed at studying the non-linear QED phenomena, such as, for instance, the Extreme Light Infrastructure, and at the existing and future lepton and hadron linear colliders.





\

We are grateful to A.~Di Piazza, A.~Chaikovskaia, A.~Tishchenko, A.~Surzhykov, A. Pupasov-Maksimov, and A.~Volotka for fruitful discussions and criticism. The studies in Sec.\,III,\,IV are supported by the Russian Science Foundation (Project No. 21-42-04412; https://rscf.ru/en/project/21-42-04412/). The studies in Sec.\,V are supported by the Ministry of Science and Higher Education of the Russian Federation (agreement No.075-15-2021-1349). The studies in Sec.\,VI are supported by the Government of the Russian Federation through the ITMO Fellowship and Professorship Program. The work on the evolved states of particles (by D. Karlovets and G. Sizykh) was supported by the Foundation for the Advancement of Theoretical Physics and Mathematics “BASIS”.

\

\begin{appendix}
\section{Appendix A: OAM of a plane wave}\label{AppA}

A plane-wave state with a definite momentum $\bp$ and helicity $\lambda$ can be expanded  over cylindrical waves as follows:
\be
|\bp, \lambda\ra = \sum\limits_{\ell = -\infty}^{\infty} i^{\ell} e^{-i \ell \phi} |p_{\perp}, p_z, m = \ell + \lambda, \lambda\ra.
\label{serl}
\ee
The probability for the plane wave to have a definite OAM projection $\la \hat{L}_z\ra = \ell = 0, \pm 1, \pm 2, ...$
\be
|i^{\ell} e^{-i \ell \phi}|^2 = 1
\label{serl}
\ee
is independent of $\ell$. This means that although the OAM mean value is vanishing for the plane wave, its OAM dispersion (or the OAM bandwidth) $\la \hat{L}_z^2\ra$ is \textit{infinitely wide}.

We can show it by explicit calculations. Let us first take the plane wave propagating strictly along the $z$ axis with the momentum $\la \bp\ra = \{0,0,\la p\ra\}$ (we explicitly write $\hbar$ in what follows),
\bea
e^{\frac{i}{\hbar}\la \bp\ra\cdot \br} = e^{\frac{i}{\hbar}\la p\ra z},
\eea
and so
\be
\hat{L}_z\, e^{\frac{i}{\hbar}\la \bp\ra\cdot \br} = \hat{L}_z^2\, e^{\frac{i}{\hbar}\la \bp\ra\cdot \br} = 0.
\ee
where
\be
\hat{L}_z = -i\hbar \frac{\partial}{\partial \phi_r}.
\ee
Such a plane wave has a definite OAM in a sense that its OAM distribution $\la \hat{L}_z^2\ra$ is vanishing together with the mean OAM value, while its azimuthal angle is undefined. 

The situation is different if the wave does \textit{not} propagate strictly along the $z$ axis and has a transverse momentum $p_{\perp}$ and a definite azimuthal angle $\phi_p$,
\be
e^{\frac{i}{\hbar}\la \bp\ra\cdot \br} = e^{\frac{i}{\hbar} p_{\perp} \rho \cos(\phi_p-\phi_r) + \frac{i}{\hbar}\la p_z\ra z},
\ee
where 
\bea
&\la \bp \ra = \{p_{\perp} \cos \phi_p, p_{\perp} \sin \phi_p, \la p_z\ra\},\cr 
&\br = \{\rho \cos \phi_r, \rho \sin \phi_r, z\}.
\eea
Let us normalize the plane wave in a large cylinder with the volume $V = L \pi R^2$, so the mean value of an operator $\hat{A}$ is
\be
\la \hat{A} \ra = \frac{\int\limits_V d^3 r\, e^{-\frac{i}{\hbar}\la \bp\ra\cdot \br} \hat{A} e^{\frac{i}{\hbar}\la \bp\ra\cdot \br}}{V = \int\limits_V d^3 r = \int\limits_{-L/2}^{L/2} dz \int\limits_{0}^{R}d\rho \rho \int\limits_0^{2\pi}d\phi}.
\ee

From here, we arrive at
\bea
&\la \hat{L}_z\ra = 0,\cr &\la \hat{L}_z^2\ra = \frac{(p_{\perp} R)^2}{4} \to \infty,\ \text{when}\ R\to \infty.
\eea
So, the definite azimuthal angle of the plane wave implies an infinitely wide OAM distribution with $\la \hat{L}_z^2\ra \to \infty$ around the central value $\la \hat{L}_z\ra = 0$.

It is exactly the situation that we encounter in the scheme \textit{(i)} of the projective measurements in the plane-wave basis in Sec.\ref{Sch}. 
Both final plane waves in a generic scattering, annihilation, or radiation process have non-vanishing transverse momenta, the definite azimuthal angles and, therefore, their mean OAMs are vanishing while the OAM dispersions are infinite.

\section{Appendix B: Working in the TAM basis}\label{Phases}

Within the plane-wave approach \textit{(i)}, the emission rates or the cross sections do not depend on the wave functions phases. Here, we are interested in phases of the evolved wave functions, which is why an analysis of the phases of the electron bispinors is in order.

\subsection{When phases matter}

A two-component spinor $w^{(\lambda)}(\theta, \phi) \equiv w^{(\lambda)}$ -- an eigenstate of the helicity operator $\hat{\bm s} \cdot \bp/|\bp|$ -- can be expanded into a series over the eigenstates of the $\hat{s}_z$ operator, 
\bea
& w^{(\lambda)} = \sum\limits_{\sigma = \pm 1/2} w^{(\sigma)}\, d^{(1/2)}_{\sigma\lambda}(\theta) e^{-i(\sigma - \lambda)\phi},\cr
& w^{(\sigma = 1/2)} = (1,0)^T,\,w^{(\sigma = -1/2)} = (0,1)^T,\cr 
& \hat{s}_z w^{(\sigma)} = \sigma w^{(\sigma)},\, \hat{j}_z w^{(\lambda)} = \lambda w^{(\lambda)},
\label{wexplambda}
\eea
and $\theta$ is the momentum polar angle, $p_{\perp} = |\bp| \sin\theta, p_z = |\bp| \cos\theta$. 
The functions
\bea
& d^{(1/2)}_{\sigma\lambda}(\theta) = \delta_{\sigma\lambda}\cos(\theta/2) - 2\sigma\,\delta_{\sigma,-\lambda} \sin(\theta/2)
\eea
are sometimes called the small Wigner functions and described, for instance, in Ref.\cite{Varsh}.

The choice of the overall phase of the spinor $w^{(\lambda)}$ defines the eigenvalue of the $\hat{j}_z$ operator. The above phase $e^{i\lambda\phi}$ corresponds to the eigenvalue $\lambda$ of the TAM operator and without this phase the eigenvalue would be vanishing, $\hat{j}_z w^{(\lambda)} = 0$.
The latter choice is made in the textbook \cite{BLP}, where the Dirac electron bispinor $\tilde{u}_{p\lambda}$, also a helicity state, looks as follows:
\bea
& \tilde{u}_{p\lambda} \equiv \sum\limits_{\sigma = \pm 1/2} u_{\varepsilon\lambda}^{(\sigma)}\, d^{(1/2)}_{\sigma\lambda}(\theta) e^{-i\sigma\phi},\cr
& u_{\varepsilon\lambda}^{(\sigma)} = \left(\sqrt{\varepsilon + m_e}\,w^{(\sigma)}, 2\lambda \sqrt{\varepsilon - m_e}\,w^{(\sigma)}\right)^T,\cr 
& \hat{s}_z u_{\varepsilon\lambda}^{(\sigma)} = \sigma u_{\varepsilon\lambda}^{(\sigma)},\ \hat{j}_z \tilde{u}_{p,\lambda} = 0,
\label{jz0}
\eea
Clearly, the operator $\hat{j}_z = \hat{s}_z + \hat{L}_z$ commutes with the Dirac Hamiltonian, so its eigenvalue is a conserved quantum number. 

The phase $e^{i\lambda \phi}$ in Eq.(\ref{wexplambda}) yields the same general phase of the bispinor $u_{p\lambda}$, which changes the orbital part and shifts the TAM to a non-vanishing value,
\bea
& \tilde{u}_{p\lambda}\, e^{i\lambda\phi} \to u_{p\lambda},\cr
& u_{p\lambda} \equiv \sum\limits_{\sigma = \pm 1/2} u_{\varepsilon\lambda}^{(\sigma)}\, d^{(1/2)}_{\sigma\lambda}(\theta) e^{-i(\sigma - \lambda)\phi},\cr
& \hat{j}_z u_{p\lambda} = \lambda u_{p\lambda}.
\label{uexplambda}
\eea
Such a choice of the phase is made in the textbook \cite{Peskin}.

The difference between the two choices is clearly seen when the $z$ axis is chosen along the momentum $\bp$, $\theta \to 0$. The state (\ref{jz0}) with $\la j_z\ra = 0$ still depends on the azimuthal angle $\phi$
\bea
w^{(\lambda)}(\theta \to 0, \phi) \to w^{(\sigma)}e^{-i\sigma\phi},
\eea
which is somewhat unnatural because it leads to a non-vanishing OAM. The state with $\la j_z\ra = \lambda$ does not have this phase factor. In what follows, we denote as $\tilde{u}_{p\lambda}$ the bispinors with a vanishing TAM projection (the choice of \cite{BLP}), while we keep $u_{p\lambda}$ for the bispinor with the TAM projection $\la\hat{j}_z\ra = \lambda$ (the choice of \cite{Peskin}). The corresponding short-hand notations are 
\be
\tilde{u}_{p\lambda} \equiv \tilde{u}\quad \text{and}\quad u_{p\lambda} \equiv u, 
\ee
respectively. Thus, whereas the phases are \textit{not} relevant during the projective measurements, for which the probability depends on $|S_{fi}^{\rm(pw)}|^2$, they can become important for generalized measurements. 

\subsection{Transition current}\label{Kinem}

When calculating matrix elements in the head-on geometry, we deal with the transition currents in momentum space. Let us calculate the current between two free fermions with the mass $m_e$, the energies $\varepsilon$ and $\varepsilon'$, and with the definite TAM $z$-projections $\lambda = \pm 1/2$ and $\lambda' = \pm 1/2$. Let the initial fermion propagate along the $z$ axis, $\bp = \{0,0,|\bp|\}$, the angles of the final fermion are $\theta', \phi'$. By using the above formulas we get
\bea
& \bar{u}_{p'\lambda'} \gamma^{\mu} u_{p\lambda} = J^{\mu}\, e^{i (\lambda - \lambda')\phi'},\cr & J^{\mu} = \{J^0, {\bm J}\},\cr
& J^{0} = d_{\lambda\lambda'}^{(1/2)}(\theta')\, \Big(\sqrt{\varepsilon + m_e} \sqrt{\varepsilon' + m_e} + 2\lambda 2\lambda' \sqrt{\varepsilon - m_e} \sqrt{\varepsilon' - m_e}\Big),\cr
& {\bm J} = \left(\sqrt{\varepsilon' + m_e}\sqrt{\varepsilon - m_e} + 2\lambda\,2\lambda'\sqrt{\varepsilon + m_e}\sqrt{\varepsilon' - m_e}\right)\cr & \times \left(d^{(1/2)}_{\lambda\lambda'}(\theta')\, {\bm \chi}_{0}  - \sqrt{2}\,d^{(1/2)}_{-\lambda\lambda'}(\theta')\, {\bm \chi}_{2\lambda}\, e^{-i 2\lambda \phi'} \right),
\label{curr}
\eea
where $\gamma^{\mu}$ are the $4\times 4$ Dirac matrices in the standard representation \cite{BLP} and
\bea
& {\bm \chi}_{0}  = (0,0,1),\;\; {\bm \chi}_{\pm 1} =\mp \fr{1}{\sqrt{2}} (1,\pm i,0),\cr
& {\bm \chi}_{\Lambda}^*\cdot{\bm \chi}_{\Lambda'} = \delta_{\Lambda\Lambda'},\cr
& \hat s_z {\bm \chi}_{\Lambda} = i\,{\bm \chi}_{0}\times {\bm \chi}_{\Lambda} = \Lambda {\bm \chi}_{\Lambda},\;\; \Lambda = 0,\pm 1.
\label{chiv}
\eea
Here, $\hat s_z$ is the photon spin operator. We have also used the equality
\bea
& (w^{(\sigma')})^{\dagger}\left\{1, \bm\sigma\right\} w^{(\sigma)} = \left\{\delta_{\sigma\sigma'}, 2\sigma \left({\bm \chi}_{0}\,\delta_{\sigma,\sigma'} - \sqrt{2}\,\delta_{\sigma,-\sigma'}\, {\bm \chi}_{2\sigma}\right)\right\},
\label{upgu}
\eea
where $\bm\sigma$ are the Pauli matrices. Note that $J^{0}$ does not depend on $\phi'$.

Likewise, we can calculate the current when the initial particle propagates opposite to the $z$ axis, $\bp = \{0,0,-|\bp|\}$. We denote the corresponding spinors of the incoming particle as $u_{p\lambda}(-\bp)$ and $\omega^{(\lambda)}(-\bp)$. We arrive at
\bea
& \bar{u}_{p'\lambda'}\gamma^{\mu}u_{p\lambda}(-\bp) = J^{\mu}(-\bp) e^{-i(\lambda + \lambda')\phi'},\cr
& J^{0}(-\bp) = d_{-\lambda\lambda'}^{(1/2)}(\theta')\, \Big(\sqrt{\varepsilon + m_e} \sqrt{\varepsilon' + m_e} + 2\lambda 2\lambda' \sqrt{\varepsilon - m_e} \sqrt{\varepsilon' - m_e}\Big),\cr
& {\bm J}(-\bp) = \left(\sqrt{\varepsilon' + m_e}\sqrt{\varepsilon - m_e} + 2\lambda\,2\lambda'\sqrt{\varepsilon + m_e}\sqrt{\varepsilon' - m_e}\right)\cr & \times \left(-d^{(1/2)}_{-\lambda\lambda'}(\theta')\, {\bm \chi}_{0} + \sqrt{2}\,d^{(1/2)}_{\lambda\lambda'}(\theta')\, {\bm \chi}_{-2\lambda}\, e^{i 2\lambda \phi'} \right).
\label{jopp}
\eea
Importantly, $u_{p\lambda}(-\bp) \ne u_{p,-\lambda}$, but $\omega^{(\lambda)}(-\bp) = \omega^{(-\lambda)}$, and the latter transition current cannot be obtained from Eq.(\ref{curr}) by simply swapping $\lambda \to -\lambda$.

Finally, in Sec.\ref{Hadrons} we encounter the 4-vector
\bea
& \sigma^{\mu\nu}q_{\nu} = - \left\{{\bm \alpha}\cdot {\bm q}, {\bm \alpha} q_0 + i {\bm \Sigma} \times {\bm q}\right\}
\eea
at the form-factor $F_2$ where all the Dirac matrices are defined as in \cite{BLP}. We need to calculate the following averages (cf. with Eq.(\ref{curr}) and Eq.(\ref{jopp})):
\bea
& \bar{u}_{p'\lambda'} {\bm \alpha} u_{p\lambda} = (2\lambda \sqrt{\varepsilon - m_p}\sqrt{\varepsilon' + m_p} - 2\lambda' \sqrt{\varepsilon + m_p} \sqrt{\varepsilon' - m_p}) \cr & \times 2\lambda \left(d_{\lambda\lambda'}^{(1/2)}(\theta') {\bm \chi}_{0} - \sqrt{2}\, d_{-\lambda\lambda'}^{(1/2)}(\theta')\, {\bm \chi}_{2\lambda}\, e^{-2i\lambda\phi'}\right) e^{i(\lambda - \lambda')\phi'},\cr
& \bar{u}_{p'\lambda'} {\bm \Sigma} u_{p\lambda} = (\sqrt{\varepsilon + m_p}\sqrt{\varepsilon' + m_p} - 2\lambda\, 2\lambda' \sqrt{\varepsilon - m_p} \sqrt{\varepsilon' - m_p}) \cr & \times 2\lambda \left(d_{\lambda\lambda'}^{(1/2)}(\theta')\, {\bm \chi}_{0} - \sqrt{2}\, d_{-\lambda\lambda'}^{(1/2)}(\theta')\, {\bm \chi}_{2\lambda}\, e^{-2i\lambda\phi'}\right)\,e^{i(\lambda - \lambda')\phi'},\cr
 & \bar{u}_{p'\lambda'} {\bm \alpha} u_{p\lambda}(-\bp) = (2\lambda \sqrt{\varepsilon - m_p}\sqrt{\varepsilon' + m_p} - 2\lambda' \sqrt{\varepsilon + m_p} \sqrt{\varepsilon' - m_p}) \cr & \times 2\lambda \left(-d_{-\lambda\lambda'}^{(1/2)}(\theta') {\bm \chi}_{0} + \sqrt{2}\, d_{\lambda\lambda'}^{(1/2)}(\theta')\, {\bm \chi}_{-2\lambda}\, e^{2i\lambda\phi'}\right) e^{-i(\lambda + \lambda')\phi'},\cr
& \bar{u}_{p'\lambda'} {\bm \Sigma} u_{p\lambda}(-\bp) = (\sqrt{\varepsilon + m_p}\sqrt{\varepsilon' + m_p} - 2\lambda\, 2\lambda' \sqrt{\varepsilon - m_p} \sqrt{\varepsilon' - m_p}) \cr & \times 2\lambda \left(-d_{-\lambda\lambda'}^{(1/2)}(\theta')\, {\bm \chi}_{0} + \sqrt{2}\, d_{\lambda\lambda'}^{(1/2)}(\theta')\, {\bm \chi}_{-2\lambda}\, e^{2i\lambda\phi'}\right)\,e^{-i(\lambda + \lambda')\phi'}.
\label{alpha}
\eea
Here, two former averages correspond to the proton moving along the $z$ axis, whereas two latter ones are calculated for the proton moving opposite to this axis. 
We have also denoted for simplicity $\phi_4 \equiv \phi', \lambda_4 \equiv \lambda', \lambda \equiv \lambda_2, \theta_4 \equiv \theta'$.

\section{Appendix C: Compton scattering and undulator radiation in the quasi-classical regime}\label{QC}


Based on the general formulas from Sec.\ref{ComptGen}, let us study in more detail the emission of soft twisted photons, $\omega' \ll \varepsilon$, by a relativistic electron, $\varepsilon \gg m_e$, which stays relativistic after the  emission, $\varepsilon' \gg m_e$. The radiated energy is concentrated at the small angles,
\begin{equation}
    \theta_{k'} \ll 1.
\end{equation}
We call this \textit{the quasi-classical regime}. Due to the delta-function $\delta (p_{\perp}' - k_{\perp}')$ in Eq.(\ref{Acylpol}),
\be
\sin\theta'/\sin\theta_{k'} = \frac{\omega'}{|\bp'|} \approx \frac{\omega'}{\varepsilon'} \ll 1,
\ee
which is why the electron scattering angle $\theta'$ is yet smaller than the photon emission angle, 
\be
\theta' \ll \theta_{k'} \ll 1.
\ee
In this case
\begin{equation}
    d^{(1/2)}_{\sigma\lambda}(\theta') = \delta_{\sigma\lambda} + O(\theta'),
    \label{dapprox}
\end{equation}
and only the term with $\sigma = -1$ survives in the sum in Eq.(\ref{Acylpol2}) in the leading approximation,
\bea
& \displaystyle {\bm A}^{(f,s)}_{(\text{g})}(\bk', \omega') \propto \sum\limits_{\sigma=0,\pm 1} \equiv \sum\limits_{\sigma=0,\pm 1} J_{s+\sigma}(\rho'_e k'_{\perp})\, e^{i (s + \sigma - \lambda')\phi_{k'}} \Big(d^{(1/2)}_{\lambda\lambda'}(\theta')\,{\bm G}_{\sigma\lambda'\lambda}^{(\uparrow \uparrow)}\, e^{i\lambda\phi_{k'}} + \cr
& \displaystyle + d^{(1/2)}_{-\lambda\lambda'}(\theta')\,{\bm G}_{\sigma\lambda'\lambda}^{(\uparrow \downarrow)}\, e^{-i\lambda\phi_{k'}}\Big) \approx \frac{(\rho'_e \omega' \theta_{k'})^{s-1}}{2^{s-1} (s-1)!}\, e^{i(s-1)\phi_{k'}}\,\left(\delta_{\lambda\lambda'}\,{\bm G}_{-1\lambda'\lambda}^{(\uparrow \uparrow)}\, + \delta_{-\lambda\lambda'}\,{\bm G}_{-1\lambda'\lambda}^{(\uparrow \downarrow)}\right)
\label{sumsigma}
\eea
where we have only taken the first term in expansion of the Bessel function in series,
\begin{equation}
    J_{s-1}(\rho'_e k'_{\perp}) \approx \frac{(\rho'_e \omega' \theta_{k'})^{s-1}}{2^{s-1} (s-1)!}.
    \label{Jassymp}
\end{equation}
In other words, the SOI vanishes in this approximation. 

Indeed, the fact that only the term with $\sigma = -1$ survives for relativistic energies is the reason why we have a minimum at $\rho = 0$ for $j_z^{(\gamma)} = 0$ in the left panel of Fig.\ref{Fig6}, whereas there is a maximum in the central panel at $\rho = 0$ for $j_z^{(\gamma)} = 1$. The Bessel function in Eq.(\ref{Acylpol3}) that defines the shape of these distributions is $J_{j_z + \sigma}(\rho p'_{\perp}) \to J_{j_z -1}(\rho p'_{\perp})$, which yields $J_1$ in the former case and $J_0$ in the latter. 

\begin{figure}[t]
	\hspace*{-0.2cm}
	\center
	\includegraphics[width=0.45\linewidth]{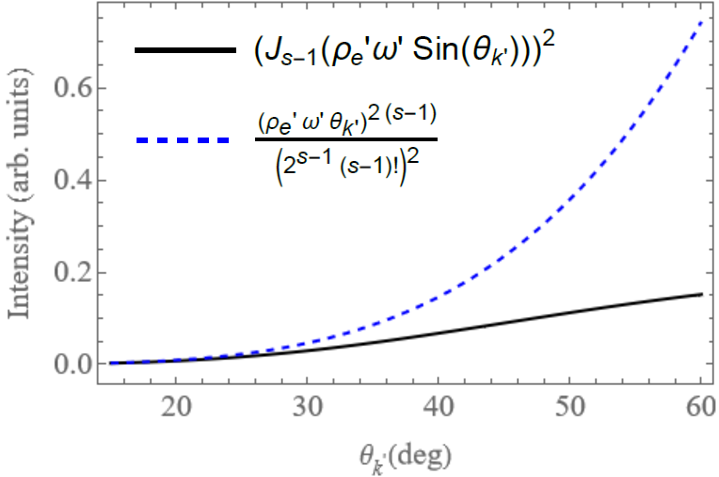}
	\caption{The angular distributions of $|{\bm A}^{(f,s)}_{(\text{g})}(\bk', \omega')|^2$ in the quasi-classical regime with no SOI, defined by $(J_{s-1}(\rho'_e \omega' \sin\theta_{k'}))^2$ according to Eq.(\ref{Acylpol2}) or by its first expansion term (\ref{Jassymp}). The latter dependence also takes place for undulator radiation at $\varepsilon \gg m_e$. Parameters: $\omega = 5\,\text{eV},\, \varepsilon/m_e = 100$ ($m_e/\varepsilon \approx 0.57$ deg), $\varepsilon' = 0.9999\, \varepsilon, \theta' \approx 10^{-5}\, \text{rad}, \eta = 0.5, s = 3$.
\label{Fig8}}
\end{figure}

As follows from the r.h.s. of Eq.(\ref{sumsigma}), the term with ${\bm G}^{(\uparrow \uparrow)}$ corresponds to the emission \textit{without an electron spin flip}, 
whereas the term with ${\bm G}^{(\uparrow \downarrow)}$ implies that the \textit{electron spin flips} when the photon is emitted.
Now let us recall that the vectors $\boldsymbol{G}^{(\uparrow \uparrow)}, \boldsymbol{G}^{(\uparrow \downarrow)}$ only depend on the electron energies 
but not on the emission angles $\theta_{k'}, \phi_{k'}$. 
As can be easily seen from Eq.(\ref{GG}), the following estimates hold in the relativistic case,
\bea
& \displaystyle f^{(1)}_{\lambda\lambda} + f^{(2)}_{\lambda\lambda} \approx 4 \sqrt{\varepsilon\varepsilon'},\cr
& \displaystyle f^{(1)}_{-\lambda\lambda} - f^{(2)}_{-\lambda\lambda} \approx 2 \sqrt{\varepsilon\varepsilon'}\, \frac{m}{\varepsilon},\cr
& \displaystyle f^{(2)}_{-\lambda\lambda} \approx \sqrt{\varepsilon\varepsilon'} \left(\frac{m}{\varepsilon'} - \frac{m}{\varepsilon}\right).
\eea
As a result
\bea
& |\boldsymbol{G}_{-1-\lambda\lambda}^{(\uparrow \downarrow)}|/|\boldsymbol{G}_{-1\lambda\lambda}^{(\uparrow \uparrow)}| = \mathcal O(m/\varepsilon),
\eea
because $\varepsilon' \lesssim \varepsilon$. In other words, the amplitude with an electron spin flip $\delta_{-\lambda\lambda'}\,{\bm G}_{-1-\lambda\lambda}^{(\uparrow \downarrow)}$ is roughly $\varepsilon/m \gg 1$ times smaller than the amplitude $\delta_{\lambda\lambda'}\,{\bm G}_{-1\lambda\lambda}^{(\uparrow \uparrow)}$ without the spin flip, which looks natural for the quasi-classical regime. As a result, only $\boldsymbol{G}_{-1\lambda\lambda}^{(\uparrow \uparrow)}$ survives and we finally get
\bea
& \displaystyle  {\bm A}^{(f,s)}_{(\text{g})} (\bk', \omega') \propto \sum\limits_{\sigma=0,\pm 1} \approx \frac{(\rho'_e \omega' \theta_{k'})^{s-1}}{2^{s-1} (s-1)!}\, e^{i(s-1)\phi_{k'}}\,\delta_{\lambda\lambda'}\,{\bm G}_{-1\lambda\lambda}^{(\uparrow \uparrow)},\cr
& \displaystyle {\bm G}_{-1\lambda\lambda}^{(\uparrow \uparrow)} \approx 2\sqrt{2}\, \eta m_e \omega\, \sqrt{\varepsilon\varepsilon'}\, {\bm \chi}_{+ 1}\, \left(\delta_{\lambda, 1/2} \left(\frac{1}{(p'k)} - \frac{1}{(pk)}\right) + \frac{1}{(pk)}\right).
\label{sumsigmafin}
\eea
When the electron recoil is completely neglected (the classical limit or the Thomson scattering), we have 
\be
\varepsilon' \to \varepsilon,\, \omega' \to s\,\omega, 
\ee
and so
\bea
& \displaystyle {\bm G}_{-1\lambda\lambda}^{(\uparrow \uparrow)} \approx 2\sqrt{2}\, \rho_e \omega\, \varepsilon\, {\bm \chi}_{+ 1},\cr
& \displaystyle \sum\limits_{\sigma=0,\pm 1} \approx \cr
& \displaystyle \approx 2\sqrt{2}\, \varepsilon\, (\rho_e \omega)^s\, \frac{\left(s\, \theta_{k'}\right)^{s-1}}{2^{s-1} (s-1)!}\, e^{i(s-1)\phi_{k'}}\,\delta_{\lambda\lambda'}\, {\bm \chi}_{+ 1}
\label{sumsigmafinclass}
\eea
where $\rho_e = ea/(pk) = \rho'_e = ea/(p'k)$ is the classical radius of the electron helical path in the plane wave from Eq.(\ref{rhoe}); see Ref.\cite{LL2}.
In Fig.\ref{Fig8} we present comparison of the general angular dependence of $|{\bm A}^{(f,s)}_{(\text{g})}(\bk', \omega')|^2$, mostly defined by the Bessel function with $\sigma = -1$, versus the paraxial behaviour (\ref{sumsigmafinclass}). The difference is only at the angles $\theta_{k'} \gg m_e/\varepsilon$.

With these approximations, we have
\bea
& \hat{j}_z{\bm A}^{(f,s)}_{(\text{g})} = s {\bm A}^{(f,s)}_{(\text{g})}.
\label{jz}
\eea
The angular distribution and the phase in Eq.(\ref{sumsigmafinclass}) \textit{exactly coincide with those of the far-field undulator radiation} (see Sec.\ref{AppB} below),
whereas the TAM (\ref{jz}) is in accord with the classical calculations for the undulator radiation and for the non-linear Thomson scattering \cite{Sasaki1, Sasaki2, Afan, Taira, Katoh, Katoh2} illustrating the correspondence principle.

Thus, not only the emitted energy distributions $|{\bm E}^{(f,s)}_{(\text{g})}|^2$ are nearly the same for Compton scattering and for emission in a helical undulator in the quasi-classical regime \cite{UndCompt}, but \textit{the phases of the fields do also coincide} in this approximation. Importantly, however, the phase $e^{i(s-1)\phi_{k'}}$ itself does not guarantee that the field carries angular momentum because the overall factor with the delta-function also plays a role. For instance, if the latter factor contains $\delta (\phi_{k}'- (\phi' \pm \pi))$ within the conventional plane--wave approach, the TAM $z$-projection is vanishing, regardless of the phase. It is only the generalized-measurement scheme with a finite error that provides vorticity of the final photons because the latter turn out to be cylindrical rather than plane waves. 

\section{Appendix D: Undulator radiation by a single electron in classical electrodynamics}\label{AppB}

In classical electrodynamics within the paraxial approximation, the inhomogeneous wave equation for the slowly-varying amplitude of the electric field in the frequency domain $\bm{\widetilde{E}}_\bot$ can be written as follows (the Gaussian units with $c=1$ are used):
 
\begin{eqnarray} \mathcal{D}
\left[\bm{\widetilde{E}}_\bot(z,\bm{r}_\bot,\omega)\right] =
\bm{g}_\bot(z, \bm{r}_\bot,\omega) ~,\label{field1}
\end{eqnarray}
where the differential operator $\mathcal{D}$ is defined by
\be
\mathcal{D} \equiv \left({\nabla_\bot}^2 + {2 i \omega }
{\partial\over{\partial z}}\right), 
\ee
${\nabla_\bot}^2$ being the Laplacian operator over transverse Cartesian coordinates.  The source-term vector  can be written in terms of the Fourier transform of
the transverse current density, $\bm{\bar{j}}_\bot(z,\bm{r}_\bot,\omega)$, and of the charge density, $\bar{\rho}(z,\bm{r}_\bot,\omega)$ (both being macroscopic quantities treated as given), as 
\bea
\bm{g}_\bot =  - {4 \pi}  \exp\left[-i \omega
	z\right] \left(i\omega\, \bm{\bar{j}}_\bot
-\bm{\nabla}_\bot \bar{\rho}\right). 
\eea
For a helical undulator, we set a constrained electron motion 
\be
\bm{r}_{o\bot}(z)=r_{ox} \bm{e}_x+r_{oy} \bm{e}_y
\ee 
with
\begin{eqnarray}
&&r_{ox}(z) = \frac{ K }{\gamma_o k_w} \cos(k_w z),\
r_{oy}(z) = \frac{ K }{\gamma_o k_w} \sin(k_w z) ~. \label{rhel0h}
\end{eqnarray}
with the constant longitudinal speed along the $z$ axis. Here
\bea
k_w = \frac{2\pi}{\lambda_w},\ K = \fr{e H_w}{k_w m_e},
\eea
$\lambda_w$ is an undulator period, and $H_w$ is the maximal modulus of the undulator magnetic field on-axis. The classical parameter $K$ is analogous to $\eta$ from Eq.(\ref{Eta}).

Solution of the wave equation is found to be:
\begin{eqnarray}
& \displaystyle \bm{\widetilde{E}}_\bot(z, \bm{r}_\bot, \omega )=
\int_{-\infty}^{z} dz' \frac{1}{z-z'} \int d
\bm{r'}_\bot \exp\left\{i\omega\left[\frac{\mid \bm{r}_\bot-\bm{r'}_\bot
	\mid^2}{2 (z-z')}\right]+ i \left[\int_{0}^{z'} d \bar{z} \frac{
	\omega }{2 {\gamma}_z^2(\bar{z}) }\right] \right\} \cr 
	& \displaystyle \times \left[\left(i\omega \bm{v}_{o\bot}(z')
-\bm{\nabla}'_\bot \right) \tilde{\rho}(z',\bm{r'}_\bot-\bm{r}_{o\bot}(z'),\omega)\right]
 ~,\cr && \label{blob}
\end{eqnarray}
%
Here $\gamma_z(z) = 1/\sqrt{1-v_{oz}(z)^2}$, and $\bm{\nabla}'_\bot$ represents the gradient operator with respect to the source point, while $(z, \bm{r}_\bot)$ indicates the observation point. The further calculations are very similar to those for a planar undulator presented in detail in Ref.\cite{Planar}.

Integration by parts of the gradient term, and introduction of a new integration variable $\bm{l}=\bm{r'}_\bot-\bm{r}_{o\bot}(z')$ gives the following expression in the far-zone limit for relativistic energies, $\gamma \gg 1$, and for the small emission angles $\theta_x = x/z \ll 1, \theta_y = y/z \ll 1$:
\begin{eqnarray}
\bm{\widetilde{E}}_\bot(z, \bm{r}_\bot, \omega ) &=& -\frac{i \omega }{z} \int
d\bm{l} \int_{-\infty}^{\infty} dz' \tilde{\rho}(z',\bm{l},\omega) {\exp{\left[i \Phi_T(z',\bm{l},\omega)\right]}} \cr && \times
\left[\left(\frac{K}{\gamma} \sin\left(k_w
z'\right)+\theta_x\right)\bm{e}_x
+\left(-\frac{K}{\gamma} \cos\left(k_w
z'\right)+ \theta_y \right)\bm{e}_y\right] \label{unduradh}
\end{eqnarray}
%
where
\begin{eqnarray}
& \displaystyle \Phi_T = {\omega} \Big\{\frac{z'}{2\gamma^2}
\Big[1+K^2 +
\gamma^2\left(\theta_x^2+\theta_y^2\right)\Big]- \cr 
& \displaystyle  - \frac{K
	\theta_x}{\gamma k_w} \cos{(k_w z')}- \frac{K
	\theta_y}{\gamma k_w} \sin{(k_w z')}\Big\} + \Phi_0
	\label{phitunduh}
\end{eqnarray}
with $\Phi_0 = \omega\left[
- (\theta_x l_x+\theta_y
l_y)+\frac{z}{2}(\theta_x^2+\theta_y^2) \right]$. Further use of the resonant approximation leads to
\bea
& \displaystyle \bm{\widetilde{E}}_{\bot s} = \frac{ K}{2\gamma}\frac{\omega_{s0}  }{z}
\int_{-\infty}^{\infty} d l_x \int_{-\infty}^{\infty} d l_y
\int_{-\infty}^{\infty} dz' \tilde{\rho}(z',\bm{l},\omega) \exp[i\Phi_0] \exp\left\{i s\frac{\Delta
	\omega_s}{\omega_s} k_w z'\right\}  \cr 
	&\displaystyle \times \frac{1}{(s-1)!} \left(s \frac{ K  \gamma}{1+K^2}\right)^{s-1}\left(\theta_y-i\theta_x\right)^{s - 1} ( \bm{e}_x + i \bm{e}_y ) . \label{undurad3h}
\eea
%
where $s$ is the harmonic number (cf. with Sec.\ref{Compt}), $\omega = \omega_s + \Delta \omega_s$. A model for \textit{the single-electron emission} is obtained by setting 
\be 
\tilde{\rho}({z},\bm{l},\omega) =
  g_{0}\left(\bm{l}\right)  \bar{f}(\omega),
\ee
  for $z$ in the range $(-L_w/2,L_w/2)$ and zero otherwise, $L_w$ is the undulator length, and $g_0 = -e \delta{(\bm{l})}$ and $ f(t) = \delta(t-t_e) \longrightarrow \bar{f}(\omega) = \exp(i \omega t_e)$. The direct substitution explicitly gives 
\bea
& \displaystyle \bm{\widetilde{E}}_{\bot s} = -e \frac{K}{2\gamma}\frac{\omega}{z}
\exp\left[i\frac{\omega_{s0} z}{2}(\theta_x^2+\theta_y^2)\right] L_w\, \mathrm{sinc}\left[C_s L_w/z +(\theta_x^2+\theta_y^2) \omega_{s0} L_w/(4)\right]  \cr & \displaystyle \frac{1}{(s-1)!} \left(s \frac{ K  \gamma}{1+K^2}\right)^{s-1}\left(\theta_y-i\theta_x\right)^{s - 1} ( \bm{e}_x + i \bm{e}_y ) . \label{undurad3h}
\eea
where  we have ignored the unimportant phase $\omega t_e$. $C_s$ is the so-called detuning parameter,
\bea
& \displaystyle C_s = \fr{\omega}{2\bar{\gamma}_z^2} - s k_w = s\,\fr{\omega - \omega_{s0}}{\omega_{s0}}\, k_w,\cr
& \displaystyle \omega_{s0} = 2s\, k_w \bar{\gamma}_z^2,\ \bar{\gamma}_z^2 = \fr{\gamma^2}{1 + K^2}.
\eea
Finally, we note that $\bm{e}_x + i \bm{e}_y = -\sqrt{2}\, \bm{\chi}_{+1}$ with $\bm{\chi}_{+1}$ from Eq.(\ref{chiv}). Going to cylindrical coordinates $\{\theta_x, \theta_y\} = \theta\{\cos\phi,\sin\phi\}$ and an infinitely long undulator, $L_w \to \infty$, we obtain in Eq.(\ref{undurad3h}) the same angular dependence and the same phase as for the non-linear Compton (or rather Thomson) scattering in the quasi-classical regime with the circularly polarized laser wave; see Eq.(\ref{sumsigmafinclass}) in Sec.\ref{QC} where $\theta_{k'}, \phi_{k'}$ are analogous to $\theta, \phi$, respectively.

\end{appendix}


\end{document}